\documentclass[11pt, draftclsnofoot, onecolumn, letterpaper]{IEEEtran}
\usepackage{graphicx}
\usepackage{amsmath}
\usepackage{amssymb}
\usepackage{amsthm}
\usepackage{mathrsfs}
\usepackage{psfrag}
\usepackage{subfigure}
\usepackage{tikz,pgfplots}

\usepackage{xspace}
\usepackage{bbm}
\input{mathlig}

\newcommand{\muspace}{\mspace{1mu}}

\DeclareRobustCommand{\scond}{\mathchoice{\muspace\vert\muspace}{\vert}{\vert}{\vert}}
\mathlig{|}{\scond}

\DeclareRobustCommand{\discint}{\mathchoice{\mspace{-1.5mu}:\mspace{-1.5mu}}{\mspace{-1.5mu}:\mspace{-1.5mu}}{:}{:}}
\mathlig{::}{\discint}
\newcommand{\suchthat}{\mathchoice{\colon}{\colon}{:\mspace{1mu}}{:}}

\def\var{\mathop{\rm Var}\nolimits}%
\def\Var{\mathop{\rm Var}\nolimits}%
\newcommand{\Ac}{\mathcal{A}}

\newcommand{\Dc}{\mathcal{D}}
\newcommand{\Ec}{\mathcal{E}}

\newcommand{\Nc}{\mathcal{N}}

\newcommand{\Sc}{\mathcal{S}}
\newcommand{\Tc}{\mathcal{T}}

\newcommand{\Xc}{\mathcal{X}}
\newcommand{\Yc}{\mathcal{Y}}

\newcommand{\Lv}{{\bf L}}

\newcommand{\lv}{{\bf l}}
\newcommand{\mv}{{\bf m}}

\newcommand{\onev}{{\boldsymbol 1}}

\newcommand{\pen}{{P_e^{(n)}}}

\newcommand{\aep}{{\mathcal{T}_{\epsilon}^{(n)}}}
\newcommand{\aepvar}{{\mathcal{T}_{\epsilon'}^{(n)}}}

\newcommand{\Mh}{{\hat{M}}}
\newcommand{\Rh}{{\hat{R}}}

\newcommand{\Xh}{{\hat{X}}}
\newcommand{\Yh}{{\hat{Y}}}
\newcommand{\Zh}{{\hat{Z}}}

\newcommand{\lh}{{\hat{l}}}
\newcommand{\mh}{{\hat{m}}}

\newcommand{\yh}{{\hat{y}}}

\newcommand{\It}{{\tilde{I}}}
\newcommand{\Jt}{{\tilde{J}}}

\newcommand{\Qt}{{\tilde{Q}}}
\newcommand{\Rt}{{\tilde{R}}}

\newcommand{\Ut}{{\tilde{U}}}

\newcommand{\Xt}{{\tilde{X}}}
\newcommand{\Yt}{{\tilde{Y}}}

\newcommand{\lt}{{\tilde{l}}}

\newcommand{\ut}{{\tilde{u}}}

\newcommand{\xt}{{\tilde{x}}}

\def\a{\alpha}
\def\b{\beta}

\def\d{\delta}
\def\e{\epsilon}

\def\th{\theta}

\DeclareMathOperator\E{\textsf{E}}
\let\P\relax
\DeclareMathOperator\P{\textsf{P}}

\DeclareMathOperator\C{\textsf{C}}

\def\error{\mathrm{e}}

\newcommand{\N}{\mathrm{N}}

\def\textiid{i.i.d.\@\xspace}
\newcommand\iid{\ifmmode\text{ i.i.d. } \else \textiid \fi}

\newcommand{\sfrac}[2]{\mbox{\small$\displaystyle\frac{#1}{#2}$}}
\newcommand{\half}{\sfrac{1}{2}}

\def\mathllap{\mathpalette\mathllapinternal}
\def\mathllapinternal#1#2{%
  \llap{$\mathsurround=0pt#1{#2}$}}

\def\clap#1{\hbox to 0pt{\hss#1\hss}}
\def\mathclap{\mathpalette\mathclapinternal}
\def\mathclapinternal#1#2{%
  \clap{$\mathsurround=0pt#1{#2}$}}

\let\oldstackrel\stackrel
\renewcommand{\stackrel}[2]{\oldstackrel{\mathclap{#1}}{#2}}

\renewcommand{\hbar}{h\mathllap{\overline{\vphantom{h}\hphantom{\rule{4.6pt}{0pt}}}\mspace{0.77mu}}}

\catcode`~=11 
\newcommand{\urltilde}{\kern -.06em\lower -.06em\hbox{~}\kern .02em}
\catcode`~=13 

\newcommand{\twov}{\mathbf{2}}
\newcommand{\elle}{l}
\newcommand{\qt}{\tilde{q}}
\newcommand{\Qh}{\hat{Q}}

\renewcommand{\th}{\hat{t}}

\newtheorem{theorem}{Theorem}
\newtheorem{lemma}{Lemma}

\newtheorem{corollary}{Corollary}

\theoremstyle{definition}
\newtheorem{example}{Example}
\newtheorem{remark}{Remark}

\begin{document}
\title{Distributed Decode--Forward for\\[-0.25em] Relay Networks}
\date{\today}
\author{Sung Hoon Lim,  Kwang Taik Kim, and Young-Han Kim%
\thanks{
This work was presented in part at
the 2014 UCSD Information Theory and Applications Workshop, La Jolla, CA;
the 2014 IEEE International Symposium of Information Theory, Honolulu, HI;
and the 2014 IEEE Information Theory Workshop, Hobart, Australia.}%
\thanks{S.~H.~Lim is with the Korea Institute of Ocean Science and Technology, Ansan, Gyeonggi-do, Korea (e-mail: shlim@kiost.ac.kr)}%
\thanks{K.~T.~Kim is with Samsung Electronics, Suwon, Gyeonggi-do, 443-742, Korea (e-mail: kwangtaik.kim@samsung.com).}%
\thanks{Y.-H.~Kim is with the Department of Electrical and Computer
Engineering, University of California, San Diego, La Jolla, CA
92093, USA (e-mail: yhk@ucsd.edu).}%
}

\maketitle
\begin{abstract}
A new coding scheme for general $N$-node relay networks is presented for unicast, multicast, and broadcast.
The proposed distributed decode--forward scheme combines and generalizes Marton coding for single-hop broadcast channels
and the Cover--El Gamal partial decode--forward coding scheme for 3-node relay channels. The key idea of the scheme is to precode all the codewords of the entire network at the source by multicoding over multiple blocks. This encoding step allows these codewords to carry partial information of the messages implicitly without complicated rate splitting and routing. This partial information is then recovered at the relay nodes and forwarded further. For $N$-node Gaussian unicast, multicast, and broadcast relay networks, the scheme achieves within $0.5N$ bits from the cutset bound and thus from the capacity (region), regardless of the network topology,
channel gains, or power constraints. Roughly speaking, distributed decode--forward is dual to noisy network coding,
which generalized compress--forward to unicast, multicast, and multiple access relay networks.
\end{abstract}

\section{Introduction}
Since van der Meulen \cite{van-der-Meulen1971a} studied the 3-node relay channel model
in the context of mathematical communication theory, numerous relaying schemes have been proposed in the literature.
Among these, decode--forward~\cite[Th.~1]{Cover--El-Gamal1979}, compress--forward~\cite[Th.~6]{Cover--El-Gamal1979},
and amplify--forward~\cite{Schein--Gallager2000, Schein2001} are particularly well studied and
form a basis for other schemes.
With different principles (digital-to-digital, analog-to-digital, and analog-to-analog, respectively) and relative strengths
over each other, the three relaying schemes have been extended beyond the 3-node relay channel and the 4-node diamond network~\cite{Schein--Gallager2000, Schein2001} with varying degrees of generality in operation and scalability in performance. Their generality
and scalability (or the lack thereof) depend heavily on the underlying network topology and message configurations (e.g., unicast, multicast, multiple access, broadcast, and multiple unicast).

Amplify--forward can be readily applied to an arbitrary Gaussian multihop network
to transform it into a single-hop network with intersymbol interference, regardless of the message configuration. Despite its
high score in generality, amplify--forward fails to achieve scalable performance as its achievable rate can have an unbounded gap from capacity
in most cases, except for a handful of special examples (cf.~\cite{Agnihotri--Jaggi--Chen2011, Maric--Goldsmith--Medard2012, Niesen--Diggavi2013}).

Compress--forward has been extended to general noisy networks by the network compress--forward scheme~\cite{Kramer--Gastpar--Gupta2005} and the noisy network coding scheme~\cite{Yassaee--Aref2011, Lim--Kim--El-Gamal--Chung2011}, the latter of which
was motivated by the quantize--map--forward scheme~\cite{Avestimehr--Diggavi--Tse2011} for Gaussian relay networks. For the multiple access message configuration (many senders communicating their messages to a single receiver over multiple hops) and its multicast extension (now to multiple receivers, each demanding the same set of messages),
the noisy network coding scheme scores high in performance scalability. For example, noisy network coding achieves
within $0.63N$ bits of capacity for an arbitrary $N$-node Gaussian multiple access relay network, regardless of the network topology, channel gains, or power constraints.
The main drawback of noisy network coding is noise propagation---the quantization noise at each relay  accumulates over multiple hops. Moreover, it is not known whether or how noisy network coding
(or any network extension of compress--forward) can achieve scalable performance for other message configurations, most notably, broadcast and multiple unicast.

Decode--forward, with its all digital operations, does not suffer much noise propagation
as in compress--forward or amplify--forward (consider, for example, a cascade of point-to-point channels)
and is expected to score the highest in performance scalability.
For a general relay network with (single-message) unicast and multicast, it has been extended by the network
decode--forward scheme~\cite{Kramer--Gastpar--Gupta2005, Xie--Kumar2005}, which achieves the capacity when the channel is physically degraded.
The scheme, however, has been extended rarely beyond unicast or multicast,
and performs rather poorly in those few exceptions (cf.~\cite{Rankov--Wittneben2006}, \cite[Sec.~19.1.2]{El-Gamal--Kim2011}).
There are two main challenges in extending decode--forward to multiple messages in a general and scalable manner.
First, the complete message decoding requirement at all (or some) of the relays is often too stringent.
Second, when there is more than one message (multiple access, broadcast, or multiple unicast),
it is unclear how the message should be routed; in other words, which relay should be assigned to forward which message.

For the 3-node relay channel, partial decode--forward \cite[Th.~7]{Cover--El-Gamal1979}, \cite{El-Gamal--Aref1982} provides a solution to the first challenge by splitting the message into two parts and letting the relay forward one of them.
This new degree of freedom in operation, however,
makes the second challenge of who forwards what more intractable
even for unicast. Consequently, except for a few extensions
for special channel models (see \cite[Sec.~3.4 and~3.5]{Aref1980},
\cite{Chern--Ozgur2014, Li--Kim2015} for unicast examples
and \cite[Rem.~17]{Kramer--Gastpar--Gupta2005} for a broadcast example),
partial decode--forward has not been extended to general relay networks.

Our discussion thus far leads to the following two questions:
\begin{itemize}
\item[1)] Can we employ relay decoding (partial or complete), which would propagate less noise, for general networks with multiple messages?

\item[2)] How can we achieve scalable performance for
message configurations beyond multiple access, for example,
for broadcast or multiple unicast?
\end{itemize}

In this paper, we provide one and a half satisfactory answers to these
questions by developing the \emph{distributed decode--forward} coding scheme.
For the one, the distributed decode--forward scheme generalizes partial decode--forward
to arbitrary networks, answering the first question
(cf.~\cite[Open problem~18.3]{El-Gamal--Kim2011}).
For the half, the distributed decode--forward scheme generalizes Marton's coding scheme
for single-hop broadcast channels to multihop broadcast relay networks. In particular,
the scheme achieves the capacity region of the general Gaussian broadcast relay network uniformly within
$0.5N$ bits per dimension, which refines the previous result by Kannan, Raja, and
Viswanath~\cite{Kannan--Raja--Viswanath2012}.
For graphical multicast networks, the scheme
achieves the network capacity as dictated by the network coding
theorem~\cite{Ahlswede--Cai--Li--Yeung2000}. In this sense,
the distributed decode--forward scheme unifies and extends
Marton coding, network coding, and partial decode--forward relaying
to general multihop networks.

The most immediate motivation of our work comes from the aforementioned work
by Kannan et al.~\cite{Kannan--Raja--Viswanath2012} on deterministic and
Gaussian broadcast relay networks. The approach taken in the current paper, as is
usually the case with any successor, is more general and versatile. In particular,
the distributed decode--forward scheme is a ``single-letter'' coding scheme
directly applicable to arbitrary network models and its performance has a clean
analytic expression that can be
easily compared to the cutset bound~\cite{El-Gamal1981b}, \cite[Th.~15.10.1]{Cover--Thomas2006}.

The distributed decode--forward scheme uses \emph{multicoding}
(see, for example, \cite{Marton1979}, \cite[Ch.~3]{Aref1980},
\cite{Kolte--Ozgur--Permuter2014})
as the main tool to overcome the aforementioned challenge
of extending partial decode--forward to networks, namely, the complexity of coordination among distributed nodes. More specifically, the source node encodes all the messages
with compatible codewords and {\it a priori} controls the transmission over the entire network.
These compatible codewords are chosen, however, via multicoding,
which allows the codewords to carry information of some part of the messages \emph{implicitly}.
As a side note, the coding scheme by Kannan et al.~\cite{Kannan--Raja--Viswanath2012}
also employs multicoding at the source and decoding at relays (at a multiletter level),
despite the prima facie observation that the essence of their scheme is quantization (compress--forward),
not decoding (decode--forward).

The rest of the paper is organized as follows. In the next section, we formally define the problem and present the main results.
In Sections~\ref{sec:unicast} and~\ref{sec:broadcast}, we develop
and analyze the distributed decode--forward scheme
for unicast and broadcast, respectively.
The results on Gaussian networks are established in Section~\ref{sec:gaussian}.
Comparison with the noisy network coding scheme is discussed
in Section~\ref{sec:discussion}, which is followed
by some concluding remarks
in Section~\ref{sec:conclusion}.

Throughout the paper, we use the notation in \cite{El-Gamal--Kim2011}. In particular, a sequence of random variables with node index $k$ and time index $i \in [1:n]:=\{1,\ldots, n\}$ is denoted by $X_k^n:=(X_{k1},\ldots, X_{kn})$. A tuple of random variables is denoted by $X(\Ac):= (X_k: k\in \Ac)$.
For a set $\Ac \in [1::N]$, the complement $\Ac^c$ is taken with respect to $[1::N]$.
Given a set of nodes $\Sc \subseteq [1::N]$, we often use the notation
\[
\Sc_k = \Sc \cap [1::k-1]\quad\text{and}\quad \Sc_k^c = (\Sc^c)_k = \Sc^c \cap [1::k-1].
\]


\section{Problem Setup and the Main Results}
\begin{figure}[b]
\begin{center}
\small
\psfrag{p1}[c]{$p(y_1,\ldots,y_N|x_1,\ldots,x_N)$}
\psfrag{m1}[r]{$(M_2,\ldots, M_N)$}
\psfrag{n1}[c]{$1$}
\psfrag{n2}[c]{$2$}
\psfrag{n3}[c]{$k$}
\psfrag{n4}[c]{$N$}
\psfrag{n5}[c]{$j$}
\psfrag{n6}[c]{$3$}
\psfrag{m2}[c]{$\Mh_2$}
\psfrag{m3}[c]{$\Mh_k$}
\psfrag{m4}[c]{$\Mh_N$}
\psfrag{m5}[c]{$\Mh_j$}
\psfrag{m6}[c]{$\Mh_3$}
\includegraphics[width=0.5\textwidth]{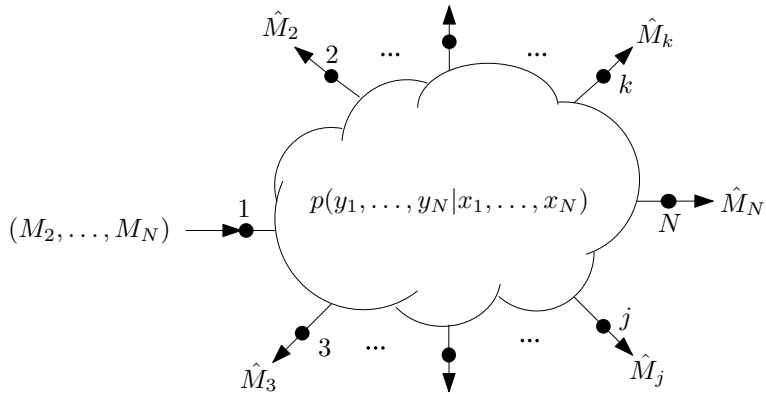}%
\caption{The $N$-node discrete memoryless broadcast network.}\label{fig:brn}
\end{center}
\end{figure}

The $N$-node discrete memoryless network (DMN) $(\Xc_1\times\cdots\times \Xc_N, p(y^N|x^N), \Yc_1\times\cdots\times \Yc_N)$ consists of $N$ sender--receiver alphabet pairs
$(\Xc_k, \Yc_k)$, $k \in [1:N]$, and a collection of conditional pmfs $p(y_1,\ldots,y_N| x_1,\ldots,x_N)$.
The noise, interference, and broadcast effects in the communication as well as the network topology (that is,
which nodes can communicate directly to which other nodes)
are defined through the structure of this conditional pmf $p(y_1,\ldots,y_N| x_1,\ldots,x_N)$.

Suppose that a single source node (node~1) wishes to communicate messages $M(\Dc)=(M_k: k\in\Dc)$
over the DMN $p(y^N | x^N)$, where message $M_k$ is intended to be recovered at node $k$ and $\Dc\subseteq [2::N]$ is the set of destination nodes,
as depicted in Fig.~\ref{fig:brn}.
Throughout this paper, we refer to this setting
as the discrete memoryless broadcast relay network (DM-BRN).
The practical motivation of this model arises from downlink communication
for cloud radio access networks \cite{china-mobile2011, Park--Simeone--Sahin--Shamai2013a, Park--Simeone--Sahin--Shamai2013b, Zhou--Yu2014}
and for distributed antenna systems with joint processing~\cite{LTEcomp2013}.

The $((2^{nR_k}:k\in\Dc), n)$ code for the DM-BRN consists of
\begin{itemize}
\item
$|\Dc|$ message sets $[1::2^{nR_k}]$, $k\in\Dc$,

\item
a source encoder
that assigns a symbol $x_{1i}(m(\Dc),y_1^{i-1})$ to each
message tuple $m(\Dc)=(m_k\in [1::2^{nR_k}]: k\in\Dc)$ and received sequence $y_1^{i-1} \in \Yc_1^{i-1}$ for $i
\in [1::n]$,

\item
a set of relay encoders, where encoder~$k \in [2::N]$ assigns $x_{ki}(y_k^{i-1})$ to each $y_k^{i-1}$ for $i \in
[1::n]$, and

\item
a set of decoders, where decoder~$k \in \Dc$ assigns an estimate
$\mh_k$ or an error message $\error$ to each $y_k^n$.
\end{itemize}
The performance of the code is measured by the average probability of error
\[
\pen = \P\{\Mh_k \ne M_k \text{ for some } k \in \Dc \},
\]
where the messages are uniformly distributed and independent of each other.
A rate tuple $(R_k: k\in\Dc)$ is said to be {\em achievable} if there exists a sequence of
$((2^{nR_k}: k\in\Dc), n)$
codes such that
$\lim_{n\to\infty} \pen  = 0$. The {\em capacity region} of the DM-BRN is the closure of the set of achievable rate tuples.

We are particularly interested in the following special cases of the general DM-BRN:
\begin{enumerate}
\item The discrete memoryless unicast relay network (DM-URN): $\Dc=\{N\}$.
\item The DM-BRN with complete destination set: $\Dc=\{2,\ldots,N\}$.
\item The discrete memoryless relay channel (DM-RC):
$N = 3$, $\Dc = \{3\}$, and $Y_1 = X_3 = \emptyset$.
\item The discrete memoryless broadcast channel (DM-BC): $Y_1 = X_2=\cdots=X_N=\emptyset$.
\item The deterministic BRN: $Y_k = g_k(X_1,\ldots, X_N)$, $k \in [1::N]$.
\item The Gaussian BRN:
\begin{align}
Y_k = g_{k1} X_1 + \cdots + g_{kN} X_N + Z_k,\quad k \in [1::N], \label{eq:gaussian-network}
\end{align}
where $g_{kj}$ is the channel gain from node~$j$ to node~$k$, and $Z_1,\ldots,Z_N$ are independent Gaussian noise components with zero mean and unit variance. We assume average power constraint $P$ on each $X_k$, i.e., $\sum_{i=1}^n \E(x^2_{1i}(m(\Dc), Y^{i-1}_1))\le nP$, $m_k\in[1::2^{nR_k}]$, $k\in\Dc$, and $\sum_{i=1}^n \E(x^2_{ki}(Y^{i-1}_k))\le nP$, $k\in[2::N]$.
\end{enumerate}

In the following, we first present the results on unicast relay networks, which are further developed
into \emph{multicast}, and then present the results on general broadcast relay networks.
In addition to pedagogical benefits, this gradual treatment allows for standalone results tailored to the unicast case
that are stronger than straightforward corollaries obtained from the results on the more general broadcast network.

\subsection{Unicast Relay Networks}

\begin{theorem}[Distributed decode--forward lower bound for unicast] \label{thm:unicast}
The capacity $C$ is lower bounded as
\begin{align}
C \ge  \max_{p(x^N, u_2^N,q)} \min_{\substack{\Sc \subseteq [1::N]:\\1 \in \Sc,\,N \in \Sc^c}}
&I(X(\Sc); U(\Sc^c), Y_N | X(\Sc^c), Q) \notag\\[-1.5em]
&\quad - \sum_{k \in \Sc^c} \bigl[I(U_k; U(\Sc_k^c), X^N | X_k, Y_k, Q) + I(X_k; X(\Sc^c_k)|Q)\bigr].  \label{eq:unicast}
\end{align}
\end{theorem}

The proof of Theorem~\ref{thm:unicast}, along with the description and analysis of the associated distributed decode--forward coding scheme, is deferred to Section~\ref{sec:unicast}.
The capacity lower bound
in~\eqref{eq:unicast} of Theorem~\ref{thm:unicast} has a similar structure to the cutset bound~\cite[Th.~15.10.1]{Cover--Thomas2006},
\begin{equation} \label{eq:cutset-unicast}
C \le \max_{p(x^N)} \min_{\substack{\Sc\subseteq[1::N]:\\1\in\Sc, N\in\Sc^c}}I(X(\Sc); Y(\Sc^c)| X(\Sc^c)).
\end{equation}
Compared to \eqref{eq:cutset-unicast}, the first term of \eqref{eq:unicast} has the auxiliary random variables $U_k$ instead of $Y_k$ (except $Y_N$) and the additional term quantifies the cost of multicoding (namely, inducing dependence among codewords).

We consider a few special cases of Theorem~\ref{thm:unicast}. First,
when specialized to the DM-RC by setting $N = 3$ and $Y_1 = X_3 = U_3 = \emptyset$, Theorem~\ref{thm:unicast} recovers
the partial decode--forward lower bound \cite[Th.~7]{Cover--El-Gamal1979} (see also \cite{El-Gamal--Aref1982} and~\cite[Th.~16.3]{El-Gamal--Kim2011}).
Thus, distributed decode--forward extends partial decode--forward to networks, answering a question raised in
\cite[Open problem~18.3]{El-Gamal--Kim2011}.

\begin{corollary}[Partial decode--forward for the DM-RC] \label{cor:pdf}
The capacity of the DM-RC $p(y_2,y_3|x_1,x_2)$ is lower bounded as
\begin{align*}
C \ge
\max_{p(x_1,x_2,u_2)} \min \bigl\{ &I(X_1, X_2; Y_3), \notag\\[-1em]
& I(X_1; U_2, Y_3 | X_2) - I(U_2; X_1 | X_2, Y_2) \bigr\}\\
= \max_{p(x_1,x_2,u_2)} \min \bigl\{ &I(X_1, X_2; Y_3), \notag\\[-1em]
& I(X_1; Y_3 | X_2, U_2) + I(U_2; Y_2 | X_2) \bigr\}.
\end{align*}
\end{corollary}

As another simple example, consider the $4$-node diamond network~\cite{Schein--Gallager2000, Schein2001}.
\begin{corollary}[Diamond network] \label{cor:diamond}
The capacity of the DM diamond network $p(y_2,y_3|x_1)p(y_4|x_2, x_3)$ is lower bounded as
{\allowdisplaybreaks
\begin{align*}
C &\ge \max_{p(x_1,x_2,x_3,u_2,u_3)} \min \bigl\{
I(X_1, X_2, X_3; Y_4), \notag\\[-1em]
& \hphantom{\;\mathrel{\ge} \max_{p(x_1,x_2,x_3,u_2,u_3)} \min \bigl\{ }
I(X_1, X_2; U_3, Y_4|X_3)-I(U_3; X_1,X_2|Y_3, X_3),\notag\\[-0.5em]
& \hphantom{\;\mathrel{\ge} \max_{p(x_1,x_2,x_3,u_2,u_3)} \min \bigl\{ }
 I(X_1, X_3; U_2, Y_4|X_2)-I(U_2; X_1,X_3|Y_2, X_2), \notag\\[-0.5em]
& \hphantom{\;\mathrel{\ge} \max_{p(x_1,x_2,x_3,u_2,u_3)} \min \bigl\{ }
I(X_1; U_2, U_3, Y_4|X_2, X_3)-I(U_2; X_1,X_3|Y_2, X_2)\notag\\[-0.5em]
& \hphantom{\;\mathrel{\ge} \max_{p(x_1,x_2,x_3,u_2,u_3)} \min \bigl\{ }
\qquad-I(U_3; X_1,X_2, U_2|Y_3, X_3)-I(X_2; X_3) \bigr\} \\
&\ge \max_{p(x_1,u_2,u_3)p(x_2,x_3)} \min \bigl\{
I(X_2, X_3; Y_4) \notag\\[-1em]
&\hphantom{\;\mathrel{\ge}\max_{p(x_1,u_2,u_3)p(x_2,x_3)} \min \bigl\{}
I(X_2; Y_4|X_3)+I(U_3; Y_3),\notag\\[-0.5em]
&\hphantom{\;\mathrel{\ge}\max_{p(x_1,u_2,u_3)p(x_2,x_3)} \min \bigl\{}
I(X_3; Y_4|X_2)+I(U_2; Y_2),\notag\\[-0.5em]
&\hphantom{\;\mathrel{\ge}\max_{p(x_1,u_2,u_3)p(x_2,x_3)} \min \bigl\{}
I(U_2; Y_2) + I(U_3;Y_3) - I(U_2; U_3) - I(X_2; X_3)
\bigr\}.
\end{align*}}
\end{corollary}


\begin{example}[Gaussian diamond relay network]
Consider the relay network
\begin{align*}
Y_2 &= g_{21} X_1 + Z_2, \\
Y_3 &= g_{31} X_1 + Z_3, \\
Y_4 &= g_{42} X_2 + g_{43} X_3 + Z_4,
\end{align*}
where the noise components $Z_k$, $k=2,3,4$, are i.i.d.\@ $\N(0,1)$ independent of $(X_1,X_2,X_3)$. The channel gain coefficients $g_{kj}$ are assumed to be real positive numbers, constant as a function of time, and known throughout the network. We assume power constraint $P$ on each sender and denote the SNR for the signal from node $k$ to node $j$ as $S_{jk}=g_{jk}^2 P$.
Suppose that in Corollary~\ref{cor:diamond} we set $X_1\sim \N(0,P)$, $U_j = g_{j1}X_1+\Zh_j$, where $\Zh_j \sim \N(0, \sigma_j^2)$ are independent of each other and of $(X_1, Y_2, Y_3)$. In addition, suppose that $(X_2, X_3)$ are zero-mean jointly Gaussian, independent of $(X_1, U_2, U_3)$, with $\E[X^2_j] = P$, $j=2,3,$  and $\E[X_2X_3]=\rho P$, $0\le \rho \le 1$.
Under this choice of the conditional distribution, the distributed decode--forward
lower bound in Corollary~\ref{cor:diamond} simplifies as
\begin{align}
C \ge \max_{\rho, \sigma_2^2, \sigma_3^2>0} \min \biggl\{ & \C(S_{42}+S_{43}+2\rho \sqrt{S_{42}S_{43}}),   \nonumber \\
& \C((1-\rho^2)S_{42})+I_3, \nonumber \\
& \C((1-\rho^2)S_{43})+I_2,  \nonumber \\
& I_2+I_3 - \half\log\frac{(\sigma_2^2+S_{21})(\sigma_3^2+S_{31})}{(\sigma_2^2\sigma_3^2+\sigma_2^2S_{31}+\sigma_3^2S_{21})(1-\rho^2)}\biggr\},
\label{eq:diamond-ddf}
\end{align}
where
\begin{align*}
I_2 &=\half\log\left(\frac{(1+S_{21})(\sigma_2^2+S_{21})}{\sigma^2_2+(1+\sigma_2^2)S_{21}}\right),\\
I_3 &=\half\log\left(\frac{(1+S_{31})(\sigma_3^2+S_{31})}{\sigma^2_3+(1+\sigma_3^2)S_{31}}\right).
\end{align*}
We compare the distributed decode--forward lower bound
with the noisy network coding lower bound~\cite{Lim--Kim--El-Gamal--Chung2011}
\begin{align}
C \ge \max_{\sigma_2^2, \sigma_3^2 > 0} \min \Bigl\{ & \C\Bigl( \sfrac{S_{21}(1+\sigma_3^2)+S_{31}(1+\sigma_2^2)}{(1+\sigma_2^2)(1+\sigma_3^2)} \Bigr),  \nonumber\\
& \C(S_{42}) + \C\Bigl( \sfrac{S_{31}}{1+\sigma_3^2}\Bigr) - \C\Bigl( \sfrac{1}{\sigma_2^2} \Bigr), \nonumber\\
& \C(S_{43}) + \C\Bigl( \sfrac{S_{21}}{1+\sigma_2^2}\Bigr) - \C\Bigl( \sfrac{1}{\sigma_3^2} \Bigr), \nonumber\\
& \C(S_{42}+S_{43}) - \C\Bigl( \sfrac{1}{\sigma_2^2}\Bigr) - \C\Bigl( \sfrac{1}{\sigma_3^2} \Bigr) \Bigr\}, \label{eq:diamond-nnc}
\end{align}
and the amplify--forward lower bound
\begin{align}
C \ge  \C\biggl( \sfrac{ \bigl( \sqrt{S_{21}S_{42}(S_{31}+1)}  + \sqrt{S_{31}S_{43}(S_{21}+1)} \bigr)^2 }{ S_{42}(S_{31}+1) + S_{43}(S_{21}+1) + (S_{21}+1)(S_{31}+1)  }  \biggr). \label{eq:diamond-af}
\end{align}
The pure decode--forward scheme achieves the lower bound
%
\begin{align}
C \ge \min \bigl\{  \C(S_{21}),\,\C(S_{31}),\, \C(S_{42}+S_{43}+2\sqrt{S_{42}S_{43}}) \bigr\}. \label{eq:diamond-df}
\end{align}
Suppose now that nodes~1 and~4 are unit distance apart, node~2 is at distance $d \in [0,1]$, and node~3 is at distance $(1-d) \in [0,1]$ from node~1 along the line between nodes~1 and~4 (see Fig.~\ref{fig:diamond_sym}). The channel gains are of the form $g_{jk} = d_{jk}^{-3/2}$, where $d_{jk}$ is the distance between nodes $j$ and $k$, hence $g_{21} =g_{43}= d^{-3/2}$, $g_{31} =g_{42} = (1-d)^{-3/2}$, and the power is $P=10$.
Fig.~\ref{fig:diamond_asym} compares the cutset bound~\cite{El-Gamal--Kim2011} on the capacity with the lower bounds achieved by distributed decode--forward~\eqref{eq:diamond-ddf}, noisy network coding~\eqref{eq:diamond-nnc}, amplify--forward~\eqref{eq:diamond-af}, and decode--forward~\eqref{eq:diamond-df}, respectively.


\begin{figure}
\begin{center}
\footnotesize
\psfrag{n1}[c]{1}
\psfrag{n2}[c]{2}
\psfrag{n3}[c]{3}
\psfrag{n4}[c]{4}
\psfrag{d1}[c]{$d$}
\psfrag{d2}[c]{$1-d$}
\psfrag{d3}[c]{1}
\hspace{6pt}\includegraphics[width=0.35\textwidth]{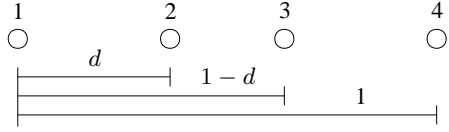}
\caption{Location of the source, relays, and destination nodes. The distance between nodes 1 and 2 is $d$, distance between nodes 1 and 3 is $1-d$, and the distance between nodes 1 and 4 is $1$.}
\label{fig:diamond_sym}
\end{center}
\end{figure}

\begin{figure}
\begin{center}
\input{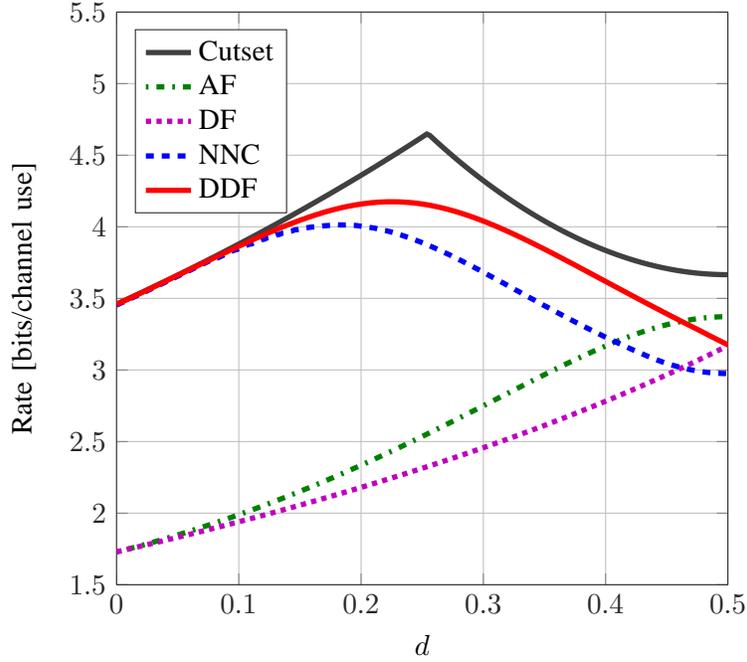}
\caption{Comparison of the cutset bound, the decode--forward lower bound (DF), the amplify--forward lower bound (AF), the noisy network coding lower bound (NNC), and the distributed decode--forward lower bound (DDF) for the Gaussian diamond network as a function of the distance $d$ where $g_{21}=g_{43}=d^{-3/2}$ and $g_{31}=g_{42}=(1-d)^{-3/2}$.}
\label{fig:diamond_asym}
\end{center}
\end{figure}
\end{example}

Next, when the channel is deterministic, we can set $U_k = Y_k$, $k \in [2::N]$, in Theorem~\ref{thm:unicast}
to establish the following.

\begin{corollary}[Deterministic URN] \label{cor:det-unicast}
If $Y_k=g_k(X_1,\ldots,X_N)$, $k \in [2::N]$, the capacity $C$ is lower bounded as
\begin{equation}
C \ge  \max_{p(x^N)}\min_{\Sc: 1\in\Sc, N\in \Sc^c} H(Y(\Sc^c) | X(\Sc^c))-\sum_{k\in\Sc^c}I(X_k; X(\Sc^c_k)).
\end{equation}
\end{corollary}
This lower bound has the same form as the cutset bound (see, for example, \cite[Sec.~18.3.1]{El-Gamal--Kim2011})
\[
C \le  \max_{p(x^N)}\min_{\Sc: 1\in\Sc, N\in \Sc^c} H(Y(\Sc^c) | X(\Sc^c))
\]
except for the negative term. With the maximum taken over pmfs of the form $p(x_2^N) = \prod_{k=2}^n p(x_k)$,
the lower bound simplifies to
\begin{equation} \label{eq:det-unicast}
C \ge  \max_{(\prod_{k=2}^Np(x_k))p(x_1|x_2^n)}\min_{\Sc: 1\in\Sc, N\in \Sc^c} H(Y(\Sc^c) | X(\Sc^c)).
\end{equation}
Accordingly, if the cutset bound is attained by a pmf of the same form,
then the lower bound in~\eqref{eq:det-unicast} is tight. For example, for graphical networks~\cite[Ch.~15]{El-Gamal--Kim2011}, Corollary~\ref{cor:det-unicast} simplifies to the max-flow min-cut theorem~\cite{Ford--Fulkerson1956}.

\smallskip

\subsection{Extension to Multicast}

Before we move on to the broadcast case, we digress briefly to discuss how our unicast results
can be generalized to the multicast setting in which the source node wishes to communicate the single
message to a set of destination nodes $\Dc\subseteq[2::N]$.
As demonstrated by Ahlswede, Cai, Li, and Yeung \cite{Ahlswede--Cai--Li--Yeung2000},
random coding allows each receiver in a multicast network to recover the common message
up to its own unicast rate. Consequently, the proof of Theorem~\ref{thm:unicast} can be adapted
in a straightforward manner to establish the following.

\begin{corollary}[Distributed decode--forward lower bound for multicast]
The capacity of the DM multicast relay network is lower bounded as
\begin{align}
C \ge \max_{p(x^N, u_2^N, q)} \min_{d\in\Dc}\min_{\substack{\Sc \subseteq [1::N]:\\1 \in \Sc,\,d \in \Sc^c}}
& I(X(\Sc); U(\Sc^c), Y_d | X(\Sc^c), Q) \notag\\[-1.5em]
&\quad- \sum_{k \in \Sc^c} \bigl[ I(U_k; U(\Sc_k^c), X^N | X_k, Y_k, Q) + I(X_k; X(\Sc^c_k)|Q)\bigr].  \label{eq:multicast}
\end{align}
\end{corollary}

Corollaries~\ref{cor:pdf} and~\ref{cor:det-unicast} can be similarly extended to the multicast case.
In particular, when specialized to graphical networks, this extension recovers the celebrated
network coding theorem \cite{Ahlswede--Cai--Li--Yeung2000},
but from a completely different path.

\subsection{Broadcast Relay Networks}

\begin{theorem}[Distributed decode--forward inner bound for broadcast] \label{thm:broadcast}
A rate tuple $(R_k \suchthat k \in \Dc)$ is achievable
for the DM-BRN $p(y^N|x^N)$ with destination set $\Dc \subseteq [2::N]$
if
\begin{align}
\sum_{k \in \Sc^c\cap\Dc} R_k < I(X(\Sc); U(\Sc^c)|X(\Sc^c), Q) -\sum_{k\in \Sc^c}\bigl[ I(U_k; U(\Sc_k^c), X^N|X_k, Y_k, Q) + I(X_k; X(\Sc^c_k)|Q)\bigr]
\label{eq:broadcast}
\end{align}
for all $\Sc\subseteq[1::N]$ such that $1 \in \Sc$ and $\Sc^c\cap\Dc\ne\emptyset$ for some $p(x^N, u_2^N, q)$.
\end{theorem}

The proof of Theorem~\ref{thm:broadcast} is deferred to Section~\ref{sec:broadcast}. As in the unicast case,
the capacity inner bound~\eqref{eq:broadcast} in Theorem~\ref{thm:broadcast} has a similar structure to the cutset bound~\cite[Th.~15.10.1]{Cover--Thomas2006} characterized by
\begin{equation} \label{eq:cutset-bc}
\sum_{k \in \Sc^c\cap\Dc} R_k \le  I(X(\Sc); Y(\Sc^c)| X(\Sc^c)),
\end{equation}
for all $\Sc\subseteq[1::N]$ such that $\Sc^c\cap\Dc\ne\emptyset$ for some $p(x^N)$.

\begin{remark}
When specialized to the unicast case, Theorem~\ref{thm:broadcast} simplifies to Theorem~\ref{thm:unicast} without $Y_N$ in~\eqref{eq:unicast}, resulting in potential rate loss.
\end{remark}

We now discuss more interesting special cases.
First, when every node $k \in [2::N]$ is a destination, i.e., $\Dc=[2::N]$, we have
$\Nc=[1::N]$, $\Sc^c \cap \Dc = \Sc^c$ and Theorem~\ref{thm:broadcast} simplifies
to the following.

\begin{corollary}[Complete destination set] \label{cor:broadcast-full}
A rate tuple $(R_k \suchthat k \in \Dc)$ is achievable
for the DM-BRN $p(y^N|x^N)$ with destination set $\Dc = [2::N]$
if
\begin{align}
\sum_{k \in \Sc^c} R_k
&< I(X(\Sc); U(\Sc^c)|X(\Sc^c), Q) \notag\\[-1em]
&\qquad-\sum_{k\in \Sc^c} \bigl[I(U_k; U(\Sc_k^c), X^N|X_k, Y_k, Q) + I(X_k; X(\Sc^c_k)|Q) \bigr]
 \label{eq:broadcast-full}
\end{align}
for all $\Sc \subseteq [1::N]$ such that $1 \in \Sc$ and $\Sc^c \ne \emptyset$
for some pmf $p(x^N, u_2^N, q)$.
\end{corollary}

Next, by setting $Y_1 = X_2=\cdots=X_N=\emptyset$ in~\eqref{eq:broadcast-full},
Corollary~\ref{cor:broadcast-full} recovers the classical result by Marton for
the single-hop broadcast channels~\cite{Marton1979}; see Appendix~\ref{app:marton} for the proof.

\begin{corollary}[Marton's inner bound for the DM-BC without common codeword]\label{cor:marton}
A rate tuple $(R_2,\ldots, R_N)$ is achievable for the DM-BC $p(y_2^N|x_1)$ if
\begin{align} \label{eq:ddf-marton}
\sum_{k \in \Sc^c} R_k  < \sum_{k\in\Sc^c} \bigl[ I(U_k; Y_k) - I(U_k; U(\Sc^c_k))\bigr]
\end{align}
for all $\Sc \subseteq [1::N]$ such that $1 \in \Sc$ and $\Sc^c \ne \emptyset$
for some pmf $p(u_2^N)$ and function $x_1(u_2^N)$.
\end{corollary}

When the channel is deterministic, we can set $U_k = Y_k$, $k \in [2::N]$, in Theorem~\ref{thm:broadcast}
to establish the following.

\begin{corollary}[Deterministic BRN] \label{cor:det-bc}
When $Y_k=g_k(X_1,\ldots,X_N)$, $k \in [2::N]$,
a rate tuple $(R_k: k\in\Dc)$ is achievable if
\begin{equation} \label{eq:deterministic}
\sum_{k \in \Sc^c \cap \Dc} R_k <  H(Y(\Sc^c) | X(\Sc^c))-\sum_{k \in \Sc^c} I(X_k; X(\Sc^c_k))
\end{equation}
for all $\Sc \subseteq [1::N]$ such that $1 \in \Sc$ and $\Sc^c \cap \Dc \ne \emptyset$
for some pmf $p(x^N)$.
\end{corollary}

This result refines
a recent result of Kannan, Raja, and Viswanath~\cite[Th.~2]{Kannan--Raja--Viswanath2012}
in which the input pmf was restricted to the form $p(x^N) = \prod_{k=1}^N p(x_k)$.
The bound in \eqref{eq:deterministic} is tight if the cutset bound is attained by the product input pmf.
Note that when specialized to graphical networks, the result by Kannan et al.~\cite[Th.~2]{Kannan--Raja--Viswanath2012} as well as the more general Corollary~\ref{cor:det-bc} recovers the broadcast capacity region
established by Federgruen and Groenevelt~\cite{Federgruen--Groenevelt1988}.


\begin{figure}[ht]
\begin{center}
\footnotesize
\psfrag{m}[b]{$M_1,M_2$}
\psfrag{x}[b]{$X_1^n$}
\psfrag{p1}[cc]{$y_2(x_1)$}
\psfrag{p2}[cc]{$y_3(x_1)$}
\psfrag{y1}[b]{$Y_2^n$}
\psfrag{y2}[b]{$Y_3^n$}
\psfrag{m1}[b]{$\Mh_2$}
\psfrag{m2}[b]{$\Mh_3$}
\psfrag{c2}[r]{$C_{23}$}
\psfrag{c3}[l]{$C_{32}$}
\psfrag{e}[c]{Encoder}
\psfrag{d1}[c]{User 2}
\psfrag{d2}[c]{User 3}
\hspace{6pt}\includegraphics[scale=0.5]{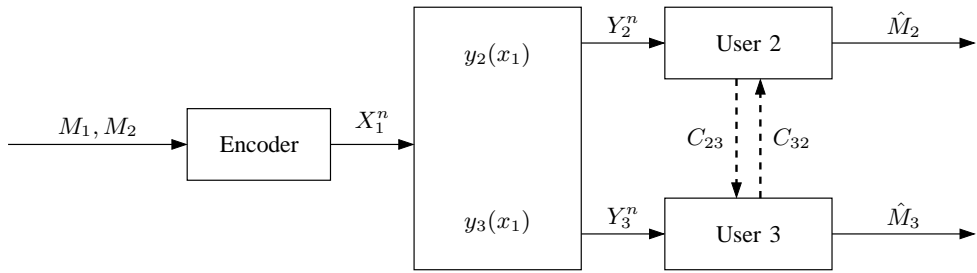}
\caption{The deterministic broadcast channel with conferencing links. }\label{fig:DBC}
\end{center}
\end{figure}

As another application of Corollary~\ref{cor:det-bc}, consider the two-user deterministic broadcast channel with cooperating receivers depicted in Fig.~\ref{fig:DBC}.
Node 1 wishes to send private messages to users 2 and 3. The source-to-destination channel is a deterministic broadcast channel $y_2(x_1), y_3(x_1)$ and
we assume that there are noiseless links from user 2 to user 3 with finite capacity $C_{23}$, and from user 3 to user 2 with finite capacity $C_{32}$. The capacity region is established by the cutset bound and a straightforward specialization of Corollary~\ref{cor:det-bc}, which consists of
all rate pairs $(R_1, R_2)$ such that
\begin{align}
R_2 &\le H(Y_2)+C_{32}, \notag\\
R_3 &\le H(Y_3)+C_{23}, \label{eq:dbc}\\
R_2+R_3 &\le H(Y_2, Y_3) \notag
\end{align}
for some $p(x_1)$.


\begin{example}[Capacity of the Blackwell channel with cooperating receivers]
\label{ex:Blackwell}
Consider a two-user deterministic broadcast channel in which the source-to-destination channel is the
Blackwell channel \cite{Gelfand1977} specified by
\begin{align*}
Y_2=\begin{cases}
0, & X_1=0 \text{ or } 2,\\
1, & X_1=1,
\end{cases}\\
Y_3=\begin{cases}
0, & X_1=0,\\
1, & X_1=1 \text{ or } 2.
\end{cases}
\end{align*}
By \eqref{eq:dbc}, the capacity region consists of all rate pairs $(R_1, R_2)$ such that
\begin{align*}
R_2 &\le H(\alpha,1-\alpha)+C_{32},\\
R_3 &\le H(\beta,1-\beta)+C_{23},\\
R_2+R_3 &\le H(\a, \b, 1-\a-\b)
\end{align*}
for some $\alpha, \beta\ge 0$ such that $\alpha+\beta\le 1$.
This result extends
the capacity region of the partially cooperating Blackwell channel (i.e., $C_{23}=0$)
investigated in \cite{Liang--Kramer2007}.
\end{example}

In~\cite{Liang--Veeravalli2007}, a general coding scheme was developed for general cooperative broadcast channels based on decode--forward and superposition coding. For the cooperative Blackwell channel in Example~\ref{ex:Blackwell}, this scheme is strictly suboptimal since superposition coding fails to achieve the capacity region of the Blackwell channel without conferencing. Thus, for this particular channel, the special case of distributed decode--forward outperforms the coding scheme in \cite{Liang--Veeravalli2007}.
This dominance is, however, not universal since neither superposition coding nor Marton coding without a common part outperforms the other when there is no conferencing ($C_{23}=C_{32}=0$).

\subsection{Gaussian Networks}

Our results hitherto on the DM-BRN can be easily adapted to
the Gaussian network model in~\eqref{eq:gaussian-network}.
In Section~\ref{sec:gaussian}, we establish the following.

\begin{theorem}[Distributed decode--forward inner bound for the Gaussian BRN] \label{thm:gaussian-bc}
A rate tuple $(R_k: k\in\Dc)$ is achievable for the Gaussian BRN if
\begin{align}
\sum_{k \in \Sc^c \cap \Dc} R_k <
\half\log |I +P G(\Sc) G^{T\!}(\Sc) | -\sum_{k\in\Sc^c}\half\log\left(1+\frac{S_k}{1+S_k}\right)
\label{eq:gaussian-broadcast}
\end{align}
for all $\Sc \subseteq [1::N]$ such that $1 \in \Sc$ and $\Sc^c \cap \Dc \ne \emptyset$,
where $S_k = \sum_{j\ne k } g_{kj}^2 P$ denotes
the received SNR at node $k$ and
the channel gain submatrix $G(\Sc)$ is defined
through
\begin{align}\label{eq:gaussian-channel}
\begin{bmatrix}
Y(\Sc) \\
Y(\Sc^c)
\end{bmatrix}
=
\begin{bmatrix}
G'\!(\Sc) & G(\Sc^c) \\
G(\Sc) & G'\!(\Sc^c)
\end{bmatrix}
\begin{bmatrix}
X(\Sc) \\
X(\Sc^c)
\end{bmatrix}
+
\begin{bmatrix}
Z(\Sc) \\
Z(\Sc^c)
\end{bmatrix}.
\end{align}
\end{theorem}

Similar results can be established for unicast (by either starting from Theorem~\ref{thm:unicast} or specializing Theorem~\ref{thm:gaussian-bc}) and multicast (by extending the unicast case).
The cutset bound for the Gaussian BRN is characterized by
\begin{align}
\sum_{k \in \Sc^c \cap \Dc} R_k \le
\half\log |I + G(\Sc) K(\Sc) G^{T\!}(\Sc) |
\label{eq:gaussian-cutset}
\end{align}
for some $K \succeq 0$ with $K_{jj} \le P$, $j \in [1::N]$,
where $K(\Sc)$ denotes the submatrix of $K$ over the indices in $\Sc$.
By comparing the cutset bound and Theorem~\ref{thm:gaussian-bc},
we establish the following gap result in Section~\ref{sec:gaussian}.

\begin{corollary}[Gaussian capacity gap] \label{thm:gaussian-bc-gap}
For the Gaussian BRN, if a rate tuple
$(R_k \suchthat k \in \Dc)$ is in the cutset bound in~\eqref{eq:gaussian-cutset},
then the rate tuple $(R_k-0.5N \suchthat k \in \Dc)$
is achievable, regardless of the channel gain matrix $G$ and power constraint $P$.
\end{corollary}

Similar gap results of $0.5N$ can be established for unicast and multicast,
which improves upon the existing
gap result of $0.63N$ achieved by noisy network coding~\cite[Th.~4]{Lim--Kim--El-Gamal--Chung2011}.
For broadcast, Corollary~\ref{thm:gaussian-bc-gap} improves upon the existing gap result of $O(N\log(N))$   by Kannan et al.~\cite[Th.~1]{Kannan--Raja--Viswanath2012}.


\section{Proof of Theorem~\ref{thm:unicast}}\label{sec:unicast}

First, we prove the achievability of
\begin{align}
R < \min_{\substack{\Sc \subseteq [1::N]:\\1 \in \Sc,\,N \in \Sc^c}}
I(X(\Sc); U(\Sc^c), Y_N | X(\Sc^c), Q)
- \sum_{k \in \Sc^c} \bigl[I(U_k; U(\Sc_k^c), X^N | X_k, Y_k, Q) + I(X_k; X(\Sc^c_k)|Q)\bigr]
\label{eq:inequality0}
\end{align}
for any pmf $p(x^N, u_2^N, q)$ such that
\begin{align}
0 \le \min_{\substack{\Sc \subseteq [1::N]:\\1 \in \Sc, \Sc^c\ne\emptyset}}
I(X(\Sc); U(\Sc^c) | X(\Sc^c), Q)
- \sum_{k \in \Sc^c} \bigl[I(U_k; U(\Sc_k^c), X^N | X_k, Y_k, Q) + I(X_k; X(\Sc^c_k)|Q)\bigr].
\label{eq:constraint0}
\end{align}
We will later show that the constraint~\eqref{eq:constraint0} on the pmf is inactive.

For simplicity, we consider the case $Q=\emptyset$. Achievability for an arbitrary $Q$ can be proved using coded time sharing~\cite[Sec.~4.5.3]{El-Gamal--Kim2011}.
We use a block Markov coding scheme in which a sequence of $(b-1)$ i.i.d.\@ messages $M_j$, $j \in [1::b-1]$, is sent over $b$ blocks
each consisting of $n$ transmissions.
For each block, we generate codewords $U_k^n$, $k \in [2::N]$, to be recovered at relay $k$. Using multicoding~\cite{Marton1979}, \cite[Sec.~7.8 and 8.3]{El-Gamal--Kim2011}, we design these codewords to be dependent among themselves and on the transmitted codewords $X_1^n, \ldots, X_N^n$. The key difference from multicoding for single-hop networks is that here multicoding is performed sequentially over multiple blocks via \emph{backward encoding}, which guarantees the desired dependence under the block Markov codebook structure.
Unlike partial decode--forward, there is no need for these codewords to have any layered superposition structure. In fact, the scheme does not keep track of which relay recovers exactly which part of the message from which node; relay $k$ recovers \emph{some} part of the message rather \emph{implicitly} by recovering $U_k^n$. The recovered part of the message, captured by an auxiliary index, is then forwarded in the next block.

We now give a detailed description of the coding scheme and provide the analysis on the probability of error.
\medskip

\noindent{\bf Codebook generation.}
Fix $p(x^N, u_2^N)$.
Randomly and independently generate a codebook for each block
$j \in [1::b]$.
Randomly and independently generate sequences $x_k^n(\elle_{k,j-1})$
according to
$\prod_{i=1}^n p_{X_k}(x_{ki})$ for $k \in [2::N]$;
sequences $x_1^n(m_j, t_j, \lv_{j-1})$
according to
$\prod_{i=1}^n p_{X_1|X_2^N}(x_{1i}|x_{2i}(\elle_{2,j-1}), \ldots, x_{Ni}(\elle_{N,j-1}))$;
and sequences $u_k^n(\elle_{kj}|\elle_{k,j-1})$
according to
$\prod_{i=1}^n p_{U_k|X_k}(u_{ki}|x_{ki}(\elle_{k,j-1}))$ for $k \in [2::N]$.
Here the values of the indices are
$m_j \in [1::2^{nR}]$,
$t_j \in [1::2^{n\Rt}]$, and
$\elle_{k,j-1}, \elle_{kj} \in[1::2^{n\Rh_k}]$, $k \in [2::N]$,
and $\lv_{j-1} := (\elle_{2,j-1}, \ldots, \elle_{N,j-1})$.
The codebooks are revealed to all parties.

Encoding and decoding are explained with the help of Table~\ref{tb:coding}.
Here the arrows indicate the direction of sequential encoding and decoding steps. For example, in the ``multicoding'' row, the arrows indicate that we start the encoding procedure for block $b$ first by finding the indices $(t_b, \lv_{b-1})$ that makes the codewords jointly typical, then move on to block $b-1$ and so forth. The $X_1$ row indicates the codewords that are sent by the source node, the $Y_k$ row indicates the recovered indices in each block, the $X_k$ row indicates the codewords that are sent by node $k$, and the $Y_N$ row indicates the recovered indices at the destination node via backward decoding.

\begin{table*}[b]
\begin{center}
\setlength{\tabcolsep}{10pt}
\small
 \begin{tabular}{c|cccccc}
\hline
\textrm{Block} & 1 & 2 & $\cdots$ & $b-1$ & $b$\\
\hline\\[-6pt]%
Multicoding &  $(t_1, \lv_{0})$ & $\leftarrow (t_2, \lv_1)$ & $\ldots$ & $\leftarrow(t_{b-1}, \lv_{b-2})$ & $\leftarrow (1,\lv_{b-1})$\\[4pt]
$X_1$ &  $x_1^n(m_1, t_1,\lv_{0})$ & $x_1^n(m_2, t_2,\lv_1)$ & $\ldots$ & $x_1^n(m_{b-1}, t_{b-1},\lv_{b-2})$ & $x_1^n(1, 1,\lv_{b-1})$\\[4pt]
$Y_k$ & $\lt_{k1} \to$ & $\lt_{k2} \to$
      & $\ldots$ & $\lt_{k,b-1} \to$
      &  1\\[4pt]
$X_k$ & $x_k^n(\lt_{k0})$ &  $x_k^n(\lt_{k1})$ & $\ldots$ & $x_k^n(\lt_{k,b-2})$
       & $x_k^n(\lt_{k,b-1})$\\[4pt]
$Y_N$ & $\mh_1$ & $\leftarrow(\mh_2, \hat\lv_1)$ & $\ldots$ & $\leftarrow(\mh_{b-1}, \hat\lv_{b-2})$ &
$\leftarrow (1,\hat\lv_{b-1})$\\[4pt]
 \hline
\end{tabular}
\end{center}
\caption{Encoding and decoding of the distributed decode--forward coding scheme for unicast.}
\label{tb:coding}
\end{table*}
\medskip

\noindent{\bf Encoding.}
The encoding procedure consists of three steps---multicoding, initialization, and actual transmission.
For $j=b,b-1,\ldots,1$, given $m_j$, find an index tuple $(t_j, \lv_{j-1})$ such that
\begin{align*}
(x_1^n(m_j, t_j, \lv_{j-1}), x_2^n(\elle_{2,j-1}), \ldots, x_N^n(\elle_{N,j-1}), u_2^n(\elle_{2j}|\elle_{2,j-1}), \ldots, u_N^n(\elle_{Nj}|\elle_{N,j-1}))\in \aepvar,
 \end{align*}
successively with the initial condition $m_b = t_b=1$ and $\lv_b=(1, \ldots, 1)$.
If there is more than one such index tuple, select one of them arbitrarily.
If there is none, let $(t_j, \lv_{j-1}) = (1,\ldots, 1)$.

Before the actual transmission of the messages,
we use additional $(N-1)^2$ blocks to transmit each $\elle_{k0}$ to node $k\in[2::N]$ using multihop coding.
The additional transmission needed for this phase is in the order of $O(nN^2)$, independent of $b$. Thus, the actual transmission rate converges to $R$ as $b\to\infty$.
In the following, we assume that all $\elle_{k0}$ indices are known prior to transmission.

To communicate the message $m_j$ in block $j$, the sender (node~$1$) transmits
$x_1^n(m_j, t_j, \lv_{j-1})$, where $t_j$ and $\lv_{j-1}$ are the indices found in the first step.

\medskip

\noindent{\bf Relay encoding.} Let $\e>\e'$. At the end of block $j$,
relay node~$k \in [2::N]$, upon receiving $y_k^n(j)$,
finds the unique index $\lt_{kj}\in[1::2^{n\Rh_k}]$ such that
\[
    (u_k^n(\lt_{kj}|\lt_{k,j-1}), x_k^n(\lt_{k,j-1}),y_k^n(j))\in \aep,
 \]
where $\lt_{k,j-1}$ is from the previous block. If there is none or more than one index, set $\lt_{kj} = 1$.
In the next block (block $j+1$),
node~$k$ transmits $x_k^n(\lt_{kj})$.
\medskip

\noindent{\bf Backward decoding.}
Decoding at the receiver (node $N$) is done backwards after all $b$ blocks are received.
For $j=b,b-1,\ldots,1$, find the unique tuple $(\mh_j, \hat \lv_{j-1})$ based on $\lh_{N,j-1} = \lt_{N,j-1}$
from relay encoding, $\hat\lv_j$ from the previous decoding step, and $y_N^n(j)$ such that
\begin{align*}
(x_1^n(\mh_j, \th_j, \hat\lv_{j-1}), x_2^n(\lh_{2,j-1}), \ldots, x_N^n(\lh_{N,j-1}), u_2^n(\lh_{2j}|\lh_{2,j-1}), \ldots, u_N^n(\lh_{Nj}|\lh_{N,j-1}), y_N^n(j))\in\aep
\end{align*}
for some $\th_j\in [1::2^{n\Rt}]$,
successively with the initial condition $\mh_b = \th_b = 1$ and $\hat \lv_b= (1,\ldots, 1)$.

\medskip

\noindent{\bf Analysis of the probability of error.}
Let $M_j$ be the message, $T_j$ and $\Lv_j$ be the indices chosen at the {\em source},
and $\tilde{\Lv}_j$ be the indices chosen at the {\em relays}.
The decoder makes an error if one or more of the following events occur:
 \begin{align*}
\Ec_0&= \bigl\{ (X_1^n(M_j, t_j, \lv_{j-1}),X_2^n(\elle_{2,j-1}), \ldots, X_N^n(\elle_{N,j-1}),
U_2^n(L_{2j}|\elle_{2,j-1}), \ldots, U_N^n(L_{Nj}|\elle_{N,j-1}) ) \not\in \aepvar \\
&\qquad\qquad \text{for all } t_j, \lv_{j-1} \text{ for some } j \in [1::b] \bigr\},\\
\Ec_1&=\bigl\{ \tilde L_{kj} \ne L_{kj} \text{ for some } k \in [2::N],\, j \in [1::b] \bigr\},\\
\Ec_2&=\bigl\{ (X_1^n(M_j, T_j, \Lv_{j-1}),X_2^n(L_{2,j-1}), \ldots, X_N^n(L_{N,j-1}),\\
 &\qquad\qquad U_2^n(L_{2j}|L_{2,j-1}), \ldots, U_N^n(L_{Nj}|L_{N,j-1}) , Y_N^n(j)) \not\in \aep \text{ for some } j \in [1::b]\bigr\},\\
\Ec_3&=\bigl\{  (X_1^n(m_j, t_j, \lv_{j-1}),X_2^n(\elle_{2,j-1}), \ldots, X_N^n(\elle_{N,j-1}), \\
&\qquad\qquad U_2^n(L_{2j}|\elle_{2,j-1}), \ldots, U_N^n(L_{Nj}|\elle_{N,j-1}), Y_N^n(j) ) \in \aep\\
&\qquad\qquad \text{for some } t_j, (m_j, \lv_{j-1}) \ne (M_j, \Lv_{j-1}), \text{ and } j \in [1::b]\bigr\}.
\end{align*}
The probability of error (averaged over the random codebook generation and messages) is upper bounded as
\begin{align}
\P(\Ec)&\le \P(\Ec_0)+\P(\Ec_1\cap \Ec_0^c)+\P(\Ec_2\cap \Ec_0^c\cap \Ec_1^c)+\P(\Ec_3). \label{eq:pe-unicast}
\end{align}

We bound each term. First, by a direct application of the properties of multivariate typicality~\cite[Sec.~2.5]{El-Gamal--Kim2011}, induction on backward encoding,
and steps similar to those of the multivariate covering lemma~\cite[Lemma~8.2]{El-Gamal--Kim2011}, the probability of $\P(\Ec_0)\to 0$ as $n\to\infty$ if
 \begin{align}
\Rt+\Rh_2+\cdots+\Rh_N &> \sum_{k=2}^N \bigl[ I(U_k; U^{k-1}, X^N|X_k) + I(X_k; X^{k-1}) \bigr]+\d(\e'),\label{eq:covering21}\\
\sum_{k\in\hat\Sc^c} \Rh_k &> \sum_{k\in\hat\Sc^c} \bigl[ I(U_k;U(\hat\Sc^c_k), X(\hat\Sc^c)|X_k)+ I(X_k; X(\hat\Sc^c_k)) \bigr]+\d(\e') \label{eq:covering22}
\end{align}
for all $\hat\Sc \subset[2::N]$, where $\hat\Sc^c = [2::N]\setminus \hat\Sc$ (here and henceforth) and $\hat\Sc^c_k = \hat\Sc^c\cap[1:k-1]$ (as in our standing notation).
The formal proof of this step is given in Appendix~\ref{sec:covering}.
By the initialization step, the conditional typicality lemma~\cite[Sec.~2.5]{El-Gamal--Kim2011},
the packing lemma~\cite[Sec.~3.2]{El-Gamal--Kim2011}, and induction on $j \in [1::b]$,
the probability $\P(\Ec_1\cap \Ec_0^c)$ tends to zero as $n\to\infty$ if
\begin{equation} \label{eq:packing1}
\Rh_k < I(U_k; Y_k| X_k)-\d(\e), \quad k\in[2::N].
\end{equation}
The probability $\P(\Ec_2\cap\Ec_0^c\cap \Ec_1^c)$ tends to zero as $n\to\infty$ due to the codebook construction
and the conditional typicality lemma.
Finally, by the union of events bound, induction on backward decoding, and the joint typicality lemma,
the last term $\P(\Ec_3)$ tends to zero as $n\to\infty$ if
\begin{align}
R+\Rt+\sum_{k \in \hat\Sc} \Rh_k &< I(X_1, X(\hat\Sc); U(\hat\Sc^c), Y_N|X(\hat\Sc^c)) \notag \\
&\quad +\sum_{k\in \hat\Sc} \bigl[ I(U_k; U(\hat\Sc_k), U(\hat\Sc^c), X^N|X_k)+I(X_k; X(\hat\Sc_k),X(\hat\Sc^c))\bigr]-\d(\e) \label{eq:packing2}
\end{align}
for all $\hat\Sc\subseteq[2::N]$ such that $N\in\hat\Sc^c$. Thus, the probability of decoding error tends to zero
as $n\to\infty$ if~\eqref{eq:covering21} through~\eqref{eq:packing2} are satisfied.
Finally, in order to obtain
the conditions on the message rate $R$ and the joint pmf $p(x^N, u_2^N)$ as in~\eqref{eq:inequality0}
and~\eqref{eq:constraint0}, respectively, we eliminate the auxiliary rates
$\Rh_2, \ldots, \Rh_k$ and $\Rt$, define $\Sc := \{1\} \cup \hat\Sc$, and take $\e \to 0$,
as shown in detail in Appendix~\ref{app:elimination}. This completes the first step of
the proof, establishing the achievability of rates satisfying~\eqref{eq:inequality0} and~\eqref{eq:constraint0}.

As the final step of the proof, we show  in Appendix~\ref{app:constraint-unicast} that the constraint~\eqref{eq:constraint0} is inactive.


\section{Proof of Theorem~\ref{thm:broadcast}}\label{sec:broadcast}

The proof of Theorem~\ref{thm:broadcast}
consists of multiple steps. First, we consider the case $\Dc = [2::N]$, which illuminates the coding scheme
at the minimal cost of notation, and establish Corollary~\ref{cor:broadcast-full}. Second, by setting
some rates to zero, we establish a capacity inner bound in~\eqref{eq:broadcast} with some constraints on the joint pmf $p(x^N,u_2^N, q)$, which can be shown to be inactive.

\subsection{Step 1: Distributed Decode--Forward for Broadcast with the Complete Destination Set}
The coding scheme for broadcast is conceptually similar to the unicast scheme,
but differs in the following three aspects. First, the messages are embedded in the auxiliary codewords $U_k^n$ instead of
the source codeword $X_1^n$. Second, $X_1^n$ now does not involve in multicoding and serves as a simple interface from the auxiliary codewords to the channel as in Marton coding~\cite{Marton1979}.
Third, the decoding step is simpler and performed in the forward direction.
We now describe the coding scheme with a sketch of its performance analysis.
For simplicity, we consider the case $Q=\emptyset$. Achievability for an arbitrary $Q$ can be proved using coded time sharing~\cite[Sec.~4.5.3]{El-Gamal--Kim2011}.
\smallskip

\noindent{\bf Codebook generation.} Fix $p(x^N, u_2^N)$. Randomly and independently generate a codebook for each block $j\in[1::b]$.
Randomly and independently generate sequences $x_k^n(\elle_{k,j-1})$
according to
$\prod_{i=1}^n p_{X_k}(x_{ki})$ for $k \in [2::N]$;
sequences $u_k^n(m_{kj}, \elle_{kj}|\elle_{k,j-1})$
according to
$\prod_{i=1}^n p_{U_k|X_k}(u_{ki}|x_{ki}(\elle_{k,j-1}))$ for $k \in [2::N]$;
and sequences $x_1^n(\mv_j, \lv_j, \lv_{j-1})$
according to
\begin{align*}
\prod_{i=1}^n p_{X_1|X_2^N, U_2^N}(x_{1i}| u_{2i}(m_{2j}, \elle_{2j}|\elle_{2,j-1}), \ldots, u_{Ni}(m_{Nj}, \elle_{Nj}|\elle_{N,j-1}), x_{2i}(\elle_{2,j-1}), \ldots, x_{Ni}(\elle_{N,j-1})).
\end{align*}
Here the values of the indices are
$m_{kj} \in [1::2^{nR_k}]$,
$\elle_{k,j-1}, \elle_{kj} \in[1::2^{n\Rh_k}]$, $k \in [2::N]$,
$\mv_j:=(m_{2j},\ldots, m_{Nj})$, and $\lv_{j} := (\elle_{2j}, \ldots, \elle_{Nj})$.
The codebooks are revealed to all parties.

Encoding and decoding are explained with the help of Table~\ref{tb:coding-bc}.

\begin{table*}[b]
\begin{center}
\setlength{\tabcolsep}{10pt}
\small
 \begin{tabular}{c|cccccc}
\textrm{Block} & 1 & 2 & $\cdots$ & $b-1$ & $b$\\
\hline\\[-6pt]%
Multicoding &  $\lv_{0}$ & $\leftarrow \lv_1$ & $\ldots$ & $\leftarrow \lv_{b-2}$ & $\leftarrow  \lv_{b-1}$\\[4pt]
$X_1$ &  $x_1^n(\mv_1, \lv_1, \lv_{0})$ & $x_1^n(\mv_2, \lv_2, \lv_1)$ & $\ldots$ & $x_1^n(\mv_{b-1}, \lv_{b-1} , \lv_{b-2})$ & $x_1^n(\mv_b, \lv_b, \lv_{b-1})$\\[4pt]
$X_k$ & $x_k^n(\lh_{k0})$ &  $x_k^n(\lh_{k1})$ & $\ldots$ & $x_k^n(\lh_{k,b-2})$
       & $x_k^n(\lh_{k,b-1})$\\[4pt]
$Y_k$ & $(\mh_{k1}, \lh_{k1})\to$ & $(\mh_{k2}, \lh_{k2})\rightarrow$ & $\ldots$ & $(\mh_{k,b-1}, \lh_{k,b-1})\rightarrow$ & $(\mh_{kb}, \lh_{kb})$\\[4pt]
 \hline
\end{tabular}
\end{center}
\vspace{-0.5em}
\caption{Encoding and decoding of the distributed decode--forward coding scheme for broadcast.}
\label{tb:coding-bc}
\vspace{-3em}
\end{table*}

\medskip

\noindent{\bf Encoding.}
The encoding procedure consists of three steps---multicoding, initialization, and actual transmission.
For $j=b,b-1,\ldots,1$, given $\mv_j$, find an index tuple $\lv_{j-1}$ such that
\begin{align*}
&(x_2^n(\elle_{2,j-1}), \ldots, x_N^n(\elle_{N,j-1}), u_2^n(m_{2j}, \elle_{2j}| \elle_{2,j-1}), \ldots,u_N^n(m_{Nj}, \elle_{Nj}| \elle_{N,j-1}))\in \aepvar,
\end{align*}
successively with the initial conditions $\mv_b=(1,\ldots,1)$ and $\lv_b=(1,\ldots,1)$.
Following similar arguments to those in Section~\ref{sec:unicast}, it can be shown that this encoding step is successful with high probability (w.h.p.)
if
\begin{equation} \label{eq:rhk}
\sum_{k \in \hat\Sc^c} \Rh_k > \sum_{k\in\hat\Sc^c} \bigl[I(U_k;U(\hat\Sc^c_k), X(\hat\Sc^c)|X_k)+I(X_k; X(\hat\Sc^c_k))\bigr]+\d(\e')
\end{equation}
for all $\hat\Sc^c\subseteq[2::N]$, where $\hat\Sc^c = [2::N]\setminus \hat\Sc$ and $\hat\Sc^c_k = \hat\Sc^c\cap[1:k-1]$.
First communicate $\lv_0$ to nodes~$2, \ldots, N$. Then communicate the message tuple $\mv_j$ in block $j$
by transmitting $x_1^n(\mv_j, \lv_j, \lv_{j-1})$,
where $(\lv_j,\lv_{j-1})$ is the index tuple chosen earlier.
\medskip

\noindent{\bf Decoding and relay encoding.}
Let $\e>\e'$. At the end of block $j = 1,\ldots, b$,
node~$k\in[2::N]$ finds the unique pair $(\mh_{kj}, \lh_{kj})$ based on $\lh_{k,j-1}$
from the previous step and $y_k^n(j)$
such that
\[
    (u_k^n(\mh_{kj}, \lh_{kj}|\lh_{k,j-1}), x_k^n(\lh_{k,j-1}), y_k^n(j))\in \aep,
\]
and declares $\mh_{kj}$ as its message estimate.
This decoding step is successful w.h.p.\@ if
\begin{equation}\label{eq:rk_rhk}
R_k+\Rh_k < I(U_k; Y_k| X_k)-\d(\e), \quad k\in[2::N].
\end{equation}
In the next block (block $j+1$), node $k$ transmits $x_{k,j+1}^n(\lh_{kj})$.

By setting $\Rh_k = I(U_k; Y_k|X_k) - R_k -2\d(\e)$ to satisfy~\eqref{eq:rk_rhk}, eliminating it from~\eqref{eq:rhk},
rewriting the conditions with $\Sc=\{1\}\cup\hat\Sc$, and taking $\e\to 0$, we have shown that any rate tuple $(R_1,\ldots, R_N)$ satisfying
\begin{align}
\sum_{k \in \Sc^c} R_k &< I(X(\Sc); U(\Sc^c)|X(\Sc^c))-\sum_{k\in \Sc^c}\bigl[I(U_k; U(\Sc_k^c), X^N|X_k, Y_k)+I(X_k; X(\Sc_k^c))\bigr] \label{eq:bcfull1}
\end{align}
for all $\Sc \subseteq [1::N]$ such that $1 \in \Sc$ and $\Sc^c \ne \emptyset$ is achievable.
By including a time sharing random variable, our argument so far is tantamount to a standalone proof of Corollary~\ref{cor:broadcast-full}.


\subsection{Step 2: Towards a General Destination Set by Projection}

Given a destination set $\Dc\subseteq[2::N]$, we set $R_k=0$, $k\not\in \Dc$, in \eqref{eq:bcfull1}.
Thus, any rate tuple $(R_k:k\in\Dc)$ is achievable  if
\begin{align}
\sum_{k \in \Sc^c\cap\Dc}R_k &< I(X(\Sc); U(\Sc^c)|X(\Sc^c), Q)-\sum_{k\in \Sc^c}
\bigl[ I(U_k; U(\Sc_k^c), X^N|X_k, Y_k, Q)+I(X_k; X(\Sc_k^c)|Q)\bigr] \label{eq:bc1}
\end{align}
for all $\Sc\subseteq[1::N]$ such that $1 \in \Sc$ and $\Sc^c\cap\Dc\ne\emptyset$
for some pmf $p(x^N, u_2^N, q)$ satisfying
\begin{align}
0 &< I(X(\Sc); U(\Sc^c)|X(\Sc^c),Q)-\sum_{k\in \Sc^c}
\bigl[I(U_k; U(\Sc_k^c), X^N|X_k, Y_k, Q)+I(X_k; X(\Sc_k^c)|Q)\bigr]  \label{eq:bc-const1}
\end{align}
for all $\Sc\subseteq[1:N]$ such that $1\in\Sc$ and $\Sc^c\cap\Dc=\emptyset$. Note that
the rate region in~\eqref{eq:bc1} is identical to the rate region in~\eqref{eq:broadcast} except
for the constraint~\eqref{eq:bc-const1} on the pmf. By continuity of mutual information,
the inequalities in~\eqref{eq:bc-const1} can be relaxed to be nonstrict.
Finally, we show
in Appendix~\ref{app:constraint-broadcast}
that the constraint~\eqref{eq:bc-const1} is inactive.
%


\section{Gaussian Broadcast Relay Networks}\label{sec:gaussian}

Theorem~\ref{thm:broadcast} for the DM-BRN can be readily extended to the Gaussian BRN
by incorporating the cost constraint and using the standard discretization method~\cite[Sec.~3.4 and~3.8]{El-Gamal--Kim2011}.
In~\eqref{eq:broadcast}, we set $X_k$, $k\in[1::N]$, to be i.i.d.\@ $\N(0,P)$, and
\begin{equation} \label{eq:aux}
U_k = g_{k1} X_1 + \cdots + g_{kN} X_N+\Zh_k, \quad k\in[2::N],
\end{equation}
where $\Zh_k \sim \N(0,1)$, $k \in [2::N]$, are mutually independent and independent of $(X^N, Y^N)$.
Note from~\eqref{eq:gaussian-channel} that $U_2^N$ and $Y_2^N$ have the same distribution and
are conditionally independent given $X^N$.
Then,
\begin{align*}
I(X(\Sc); U(\Sc^c) | X(\Sc^c))&= I(X(\Sc); Y(\Sc^c) | X(\Sc^c)) \\
&= \half\log |I +P G(\Sc) G^{T\!}(\Sc) |
\end{align*}
and
\begin{align*}
I(U_k; U(\Sc_k^c), X^N | X_k, Y_k) &= I(U_k; X^N | X_k, Y_k)\\
&= \half\log\left(1+\frac{S_k}{1+S_k}\right)\le\frac{1}{2},
\end{align*}
where $S_k = \sum_{j \ne k} g_{kj}^2 P$. Plugging these into \eqref{eq:broadcast}
establishes the inner bound in Theorem~\ref{thm:gaussian-bc}, which can be further relaxed to
\begin{align}
\sum_{k \in \Sc^c\cap\Dc} R_k < \half\log |I +P G(\Sc) G^{T\!}(\Sc) | -\frac{|\Sc^c|}{2}. \label{eq:relaxed-inner}
\end{align}

To prove Corollary~\ref{thm:gaussian-bc-gap}, we relax the cutset bound in~\eqref{eq:gaussian-cutset} as
\begin{align}
\half\log |I+G(\Sc)  K_{X(\Sc)} G^{T\!}(\Sc)  |
&\le \half\log |I +P G^{T\!}(\Sc) G(\Sc) | + \half \log \Bigl| I+\frac{1}{P}K_{X(\Sc)}\Bigr| \notag\\
&\stackrel{(a)}{\le} \half\log|I +P G(\Sc) G^{T\!}(\Sc) | + \frac{|\Sc|}{2}, \label{eq:rcs}
\end{align}
where $K_{X(\Sc)}$ is the covariance matrix of $X(\Sc)$ and
$(a)$ follows by the Hadamard inequality.
Comparing the inner and outer bounds, we can conclude that distributed decode--forward achieves within $0.5N$ bits per dimension
from the cutset bound and thus from the capacity region.

\begin{remark}
If we set $(X_1,\ldots,X_N)\sim \N(0,K)$
with $\E[X_k^2] = \rho_k^2 \le P$, $k\in[1::N]$,
and $(\Zh_2,\ldots,\Zh_N)\sim \N(0,\Sigma)$ with $\E[\Zh_k^2] = \sigma_k^2 > 0$
in~\eqref{eq:aux},
then the corresponding inner bound in Theorem~\ref{thm:gaussian-bc} is characterized by
\begin{align}
\sum_{k \in \Sc^c \cap \Dc} R_k
&< \half\log |\Sigma(\Sc^c) + G(\Sc) K(\Sc |\Sc^c) G^{T\!}(\Sc) | + \half\log|K(\Sc^c)|\notag\\
&\qquad - \sum_{k\in\Sc^c}
 \left[ \half\log\left(\sigma^2_k+\frac{S_k}{1+S_k}\right)
 + \half\log \rho_k^2 \right]  \label{eq:gaussian-broadcast-fulleval}
\end{align}
where
$K(\Sc^c)$ and $\Sigma(\Sc^c)$ are the covariance matrices of $X(\Sc^c)$ and $\Zh(\Sc^c)$, respectively,
$K(\Sc |\Sc^c)$ is the conditional covariance matrix of $X(\Sc)$ given $X(\Sc^c)$,
and
\[
S_k = \Var(Y_k - Z_k | X_k) = \Var({\textstyle \sum_{j\ne k} g_{kj} X_j}| X_k).
\]
Compared to the previous choice of $K = PI$ and $\Sigma = I$,
the improvement can be significant. For example, the gap from capacity
for the two-hop $N$-relay diamond network can be tightened from
$O(N)$ to $O(\log N)$ as reported in~\cite{Li--Kim2015}.
\end{remark}

\begin{remark}
The results in Theorem~\ref{thm:gaussian-bc} and Corollary~\ref{thm:gaussian-bc-gap}
can be readily extended to Gaussian vector (MIMO) broadcast networks.
In particular, distributed decode--forward achieves within $0.5 T$ bits from the capacity
for a Gaussian vector broadcast network with total $T$ antennas
(cf.~\cite{Courtade--Ozgur2015}).

\end{remark}

\section{Discussion}\label{sec:discussion}

\begin{figure}[t]
\begin{center}
\footnotesize
\psfrag{p1}[c]{$p(y_1,\ldots,y_N|x_1,\ldots,x_N)$}
\psfrag{m1}[r]{$(\Mh_2,\ldots, \Mh_N)$}
\psfrag{n1}[c]{$1$}
\psfrag{n2}[c]{$2$}
\psfrag{n3}[c]{$k$}
\psfrag{n4}[c]{$N$}
\psfrag{n5}[c]{$j$}
\psfrag{n6}[c]{$3$}
\psfrag{m2}[c]{$M_2$}
\psfrag{m3}[c]{$M_k$}
\psfrag{m4}[c]{$M_N$}
\psfrag{m5}[c]{$M_j$}
\psfrag{m6}[c]{$M_3$}
\includegraphics[width=0.45\textwidth]{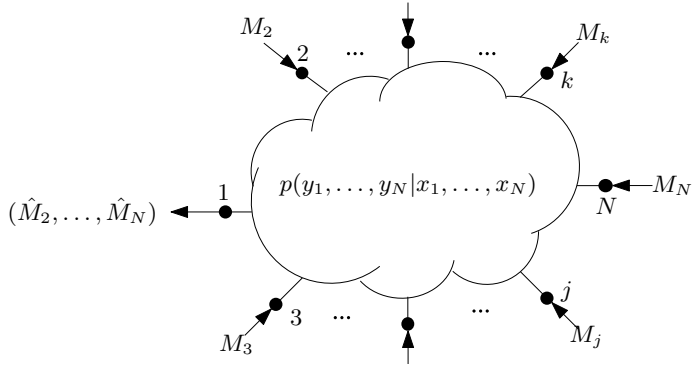}%
\caption{The $N$-node discrete memoryless multiple access network.}\label{fig:mac}
\end{center}
\end{figure}

As a dual setting to the broadcast relay network, consider
the \emph{multiple access relay network} $p(y^N | x^N)$,
in which source nodes $k\in [2::N]$
communicate independent messages to the common destination node~$1$ as depicted in Fig.~\ref{fig:mac}.
This is a special case of the multimessage multicast network~\cite[Sec.~18.4]{El-Gamal--Kim2011}
and the noisy network coding scheme \cite{Lim--Kim--El-Gamal--Chung2011, Yassaee--Aref2011, Hou--Kramer2016} yields a capacity inner bound that consists of
all rate tuples $(R_2, \ldots, R_N)$ such that
\begin{align}
\sum_{k \in \Sc} R_k &<  I(X(\Sc); \Yh (\Sc^c), Y_1|X(\Sc^c)) \notag\\[-1em]
&\qquad- I(Y(\Sc);\Yh(\Sc)|X^N, \Yh(\Sc^c), Y_1) \label{eq:mac}
\end{align}
for all $\Sc\subseteq [1::N]$ such that $1\in \Sc^c$ and $\Sc \ne \emptyset$
for some pmf $\prod_{k=1}^N p(x_k)p(\yh_k|y_k,x_k)$.
This noisy network coding inner bound in~\eqref{eq:mac} can be readily compared to the
distributed decode--forward inner bound in Corollary~\ref{cor:broadcast-full},
which is repeated here and is characterized by
\begin{align}
\sum_{k \in \Sc^c} R_k
&< I(X(\Sc); U(\Sc^c)|X(\Sc^c)) \notag\\[-1em]
&\qquad-\sum_{k\in \Sc^c} \bigl[I(U_k; U(\Sc_k^c), X^N|X_k, Y_k) + I(X_k; X(\Sc^c_k)) \bigr]
\label{eq:broadcast-full2}
\end{align}
for all $\Sc \subseteq [1::N]$ such that $1 \in \Sc$ and $\Sc^c \ne \emptyset$
for some $p(x^N, u_2^N)$.

The correspondence between two bounds is apparent.
The first term in~\eqref{eq:mac} has an auxiliary random variable $\Yh_j$, which is to be encoded at node~$j$ and to be decoded at the destination node~$1$. In comparison, the first term in~\eqref{eq:broadcast-full2}
has an auxiliary random variable $U_j$, which is to be encoded at source node~$1$ and to be decoded at node~$j$. In addition, the second term in~\eqref{eq:mac} quantifies the cost of decoding $\Yh_j$ at
the destination node, while the second term in~\eqref{eq:broadcast-full2} quantifies the cost of encoding
$U_j$ at the source node.
We highlight some of the obvious correspondences between~\eqref{eq:mac} and~\eqref{eq:broadcast-full2}
in Table~\ref{tb:comparison}.

\begin{table*}[h]
\begin{center}
\setlength{\tabcolsep}{10pt}
\normalsize
\begin{tabular}{ccc}
\hline
\rule[-9pt]{0pt}{22pt}
& Distributed decode--forward & Noisy network coding\\[-2pt]
\hline 
\rule{0pt}{12pt}
Network model & Broadcast relay networks & Multiple access relay networks\\
Single-hop & Marton inner bound & MAC capacity region \\
Relay channel & Partial decode--forward lower bound & Compress--forward lower bound \\
Gaussian capacity gap & $0.5 N $ bits & $0.63 N$ bits\\
\hline
\end{tabular}
\end{center}
\vspace{-0.5em}
\caption{Comparison between the distributed decode--forward and noisy network coding inner bounds.}
\label{tb:comparison}
\vspace{-3em}
\end{table*}

This duality between the two inner bounds is also reflected by the operations of
the coding schemes.
In noisy network coding, the sources and the relays are relatively simple, but the major burden is on the destination to recover the messages and the compression indices from the entire network.
Thus, this scheme fits well with \emph{uplink} multihop communication.
In distributed decode--forward, the relays and the destinations are relatively simple, but the source needs to precode dependent codewords for the entire network. Thus, this scheme fits well with
\emph{downlink} multihop communication.
This operational reciprocity in the roles of source and destination for multiple access and broadcast has been well noted by Kannan, Raja, and Viswanath~\cite{Kannan--Raja--Viswanath2012}, which was the key intuition for their coding scheme that parallels the quantize--map--forward scheme by Avestimehr, Diggavi, and Tse~\cite{Avestimehr--Diggavi--Tse2011}.

\section{Concluding Remarks}
\label{sec:conclusion}

Consider the $N$-node relay network $p(y^N|x^N)$ with $3^N - 2^{N+1} + 1$ messages (flows) $M_f$ between every possible pair of disjoint source--destination sets $(\Sc_f, \Dc_f)$.
This setting includes all standard channel coding problems studied in network information theory and is, in some sense, the most general network communication problem.
Finding the capacity region (optimal tradeoff between flow rates) seems to be intractable, with several results~\cite{Xiang--Wang--Kim2011, Nair--Xia--Yazdanpanah2015} in the literature hinting that
a computable expression may not exist even for simple special cases.
A natural next step is to develop a ``good'' coding scheme and, through it, provide a ``reasonable'' approximation of the capacity region.

When there are multiple flows from distinct sources to a common destination set (e.g., unicast, multicast, multiple access), noisy network coding
and the cutset bound provide a reasonable approximation of the capacity.
When there are multiple flows from a common source set to distinct destinations (e.g., unicast, multicast, broadcast), distributed decode--forward
and the cutset bound provide a reasonable approximation of the capacity. Both developments trace back to two canonical coding schemes
for the 3-node relay channel, namely, compress--forward and partial decode--forward, and their combination~\cite[Th.~7]{Cover--El-Gamal1979};
see \cite{Lee--Chung2015} for a combination of noisy network coding and distributed decode--forward for a general unicast relay network.

With these developments, the second question we asked earlier in Section~I should now read:
\begin{itemize}
\item[2$'$)] How can we achieve scalable performance for
message configurations beyond multiple access and broadcast, for example,
for broadcast with a common message and multiple unicast?
\end{itemize}

Broadcast with a common message (multiple flows to distinct destinations and a single flow to all destinations) can be viewed as the simplest combination of multicast and broadcast. Nonetheless,
the problem seems to be quite challenging even for graphical networks; see, for example,
\cite{Zou--Nosratinia--Sayed2006, Vasudevan--Korada2008}.
So far it is unclear how distributed decode--forward can be adapted to this setting.

Multiple unicast (multiple flows from distinct sources to distinct destinations)
brings up several new challenges to capacity approximation.
On the one hand, the cutset bound for multiple unicast can be significantly improved in general by the directed cutset bound~\cite{Kamath--Kim2014},
which can be still quite loose even for index coding~\cite{Blasiak--Kleinberg--Lubetzky2011,
Sun--Jafar2015}, which is a simple special case of graphical networks.
On the other hand, no classical coding schemes or their variations
have been successfully extended to multiple unicast,
a few existing results based on structured coding~\cite{Korner--Marton1979, Nazer--Gastpar2011, Nam--Chung--Lee2010, Wilson--Narayanan--Pfister--Sprintson2010} are mostly sui generis~\cite{Niesen--Nazer--Whiting2013, Ordentlich--Erez--Nazer2014} and difficult to combine with random coding schemes such as noisy network coding.
A new coding scheme is on order.

\section*{Acknowledgments}
The authors would like to thank Si-Hyeon Lee for pointing out
the possibility of using arbitrary joint pmfs in the coding scheme.
They would also like to thank Abbas El Gamal for helpful discussions.
\appendices

\section{Proof of Corollary~\ref{cor:marton}} \label{app:marton}
Consider
\begin{align*}
&I(X_1; U(\Sc^c))-\sum_{k\in\Sc^c}I(U_k; U(\Sc^c_k), X_1|Y_k)\\
&\qquad= \sum_{k\in\Sc^c} I(U_k; X_1| U(\Sc_k^c))-\sum_{k\in\Sc^c}I(U_k; U(\Sc^c_k), X_1|Y_k)\\
&\qquad\stackrel{(a)}{=} \sum_{k\in\Sc^c} \bigl[ I(U_k; X_1| U(\Sc_k^c))
 - I(U_k; U(\Sc^c_k), X_1) + I(U_k; Y_k) \bigr]\\
&\qquad= \sum_{k\in\Sc^c} \bigl[ I(U_k; Y_k)- I(U_k; U(\Sc^c_k)) \bigr],
\end{align*}
where $(a)$ follows since $U_2^N\to X_1\to Y_k$ form a Markov chain.


\section{Analysis of $\P(\Ec_0)$ in \eqref{eq:pe-unicast}}\label{sec:covering}
By the union of events bound, we have
\begin{align}
\P(\Ec_0) &\le \sum_{j=1}^b
       \P\bigl\{ (X_1^n(M_j, t_j, \lv_{j-1}),X_2^n(\elle_{2,j-1}), \ldots, X_N^n(\elle_{N,j-1}), \notag \\[-1em]
&\qquad\qquad\quad
                  U_2^n(L_{2j}|\elle_{2,j-1}), \ldots, U_N^n(L_{Nj}|\elle_{N,j-1}) ) \not\in \aepvar \text{for all } t_j, \lv_{j-1} \bigr\} \notag\\
&\le b \P\bigl\{ (X_1^n(t, \lv),(U_2^n, X_2^n)(\elle_2), \ldots, (U_N^n,X_N^n)(\elle_N)
                  ) \not\in \aepvar \text{for all } t, \lv \bigr\},
\label{eq:PE0}
\end{align}
where
$(U_k^n, X_k^n)(\elle_k)$, $\elle_k \in [1::2^{n\Rh_k}]$, are distributed independently  according to
$\prod_{i=1}^n p_{U_k,X_k}(u_{ki}, x_{ki})$, $k \in [2::N]$,
and for each $\lv = (\elle_2,\ldots,\elle_N)$, $X_1^n(t, \lv)$, $t \in [1::2^{n\Rt}]$, are
distributed conditionally independently  according to
$\prod_{i=1}^n p_{X_1|X_2^N}(x_{1i}|x_{2i}(\elle_2), \ldots, x_{Ni}(\elle_N))$.

Let $\Ac = \{(t, \lv) \suchthat (X_1^n(t, \lv),(U_2^n,X_2^n)(\elle_2), \ldots, (U_N^n,X_N^n)(\elle_N)) \in \aepvar \}$.
Then, by the Chebyshev lemma~\cite[App.~B]{El-Gamal--Kim2011},
the probability in~\eqref{eq:PE0} is upper bounded as
\begin{equation}
    \P\{|\Ac|=0\}\le \frac{\text{Var}(|\Ac|)}{(\E|\Ac|)^2}.
\end{equation}
Using indicator random variables, we express $|\Ac|$ as
\[
|\Ac| = \sum_{t, \lv} E(t, \lv),
\]
where
\[
E(t,\lv) =
    \begin{cases}
    1 &\text{ if } (X_1^n(t, \lv), (U_2^n,X_2^n)(\elle_2), \ldots, (U_N^n,X_N^n)(\elle_N)
                   ) \in \aepvar,\\
    0 &\text{ otherwise}.
    \end{cases}
\]
Denote $\lv = \onev$ if $\elle_k = 1$, $k \in [2::N]$. Similarly, for $\hat\Sc \subseteq [2::N]$, denote $\lv = \twov(\hat\Sc)$ if $\elle_k = 2$, $k \in \hat\Sc$,
and $\elle_k = 1$, $k \in \hat\Sc^c$.
Let
\begin{align*}
p_1&=\P\{E(1,\onev)=1\},\\
p_2(\hat\Sc)&= \P\{E(1,\onev)=1, E(1,\twov(\hat\Sc))=1\}, \quad \hat\Sc\subseteq[2::N],\\
p_3(\hat\Sc)&= \P\{E(1,\onev)=1, E(2,\twov(\hat\Sc))=1\}, \quad \hat\Sc\subseteq[2::N].
\end{align*}
Then
\[
\E(|\Ac|) =\sum_{t,\lv} \P\{E(t,\lv)=1\} = 2^{n(\Rt+\Rh_2^N)}p_1
\]
and
\begin{align*}
\E(|\Ac|^2)&=\sum_{t,\lv}\sum_{t',\lv'}\P\{E(t, \lv)=1, E(t',\lv')=1\}\\
&=\sum_{t,\lv} \P\{E(t, \lv)=1\}
+ \sum_{t,\lv} \sum_{\substack{\hat\Sc \subseteq [2::N]:\\\hat\Sc\ne\emptyset}}
  \sum_{\lv': l'_k \ne l_k \text{ iff } k \in \hat\Sc} \P\{E(t, \lv)=1, E(t,\lv')=1\} \\
&\quad
+ \sum_{t,\lv} \sum_{t'\ne t} \sum_{\hat\Sc \subseteq [2::N]}
  \sum_{\lv': l'_k \ne l_k \text{ iff } k \in \hat\Sc} \P\{E(t, \lv)=1, E(t',\lv')=1\} \\
&\le 2^{n(\Rt+\Rh_2^N)}p_1
+\sum_{\substack{\hat\Sc\subseteq[2::N]:\\\hat\Sc\ne\emptyset}}2^{n(\Rt+\Rh_2^N+\Rh(\hat\Sc))}p_2(\hat\Sc)
+\sum_{\hat\Sc\subseteq[2::N]}2^{n(2\Rt+\Rh_2^N+\Rh(\hat\Sc))}p_3(\hat\Sc),
\end{align*}
where $\Rh_2^N = \sum_{k=2}^N \Rh_k$ and $\Rh(\hat\Sc) = \sum_{k \in \hat\Sc} R_k$.
Since $p_3([2::N])=p_1^2$,
\begin{align*}
\var(|\Ac|)&=\E(|\Ac|^2)-(\E(|\Ac|))^2\\
&\le
2^{n(\Rt+\Rh_2^N)}p_1
+\sum_{\substack{\hat\Sc\subseteq[2::N]:\\\hat\Sc\ne\emptyset}}2^{n(\Rt+\Rh_2^N+\Rh(\hat\Sc))}p_2(\hat\Sc)
+\sum_{\substack{\hat\Sc\subseteq[2::N]:\\\hat\Sc\ne[2::N]}}2^{n(2\Rt+\Rh_2^N+\Rh(\hat\Sc))}p_3(\hat\Sc).
\end{align*}
Now it can be checked by the joint typicality lemma~\cite{El-Gamal--Kim2011} that,
for $n$ sufficiently large, we have
\begin{align*}
p_1 &\ge 2^{-n(I+\d(\e'))}, \\
p_2(\hat\Sc) &\le 2^{-n(I+J(\hat\Sc)-\d(\e'))}, \quad \hat\Sc\subseteq[2::N], \hat\Sc\ne \emptyset,\\
p_3(\hat\Sc) &\le 2^{-n(I+J(\hat\Sc)-\d(\e'))}, \quad \hat\Sc\subseteq[2::N], \hat\Sc\ne [2::N],
\end{align*}
where
\begin{align*}
I&=H(X_1|X_2^N) + \sum_{k=2}^N H(U_k, X_k)-H(X_1, X_2^N, U_2^N),\\
J(\hat\Sc)&=H(X_1|X_2^N)+\sum_{k\in\hat\Sc} H(U_k, X_k)-H(X_1, U(\hat\Sc), X(\hat\Sc)| U(\hat\Sc^c), X(\hat\Sc^c)),
\end{align*}
and $\hat{\Sc}^c = [2::N]\setminus \hat{\Sc}$.
Therefore,
\begin{align*}
\frac{\var(|\Ac|)}{(\E|\Ac|)^2} &\le 2^{-n(\Rt+\Rh_2^N-I-\d(\e'))}+\sum_{\substack{\hat\Sc\subseteq[2::N]:\\\hat\Sc\ne\emptyset}}2^{-n(\Rt+\Rh(\hat\Sc^c)-I+J(\hat\Sc^c)-3\d(\e'))} \\
&\qquad
 + \sum_{\substack{\hat\Sc\subseteq[2::N]:\\\hat\Sc\ne[2::N]}}2^{-n(\Rh(\hat\Sc^c)-I+J(\hat\Sc^c)-3\d(\e'))},
\end{align*}
which tends to zero as $n\to\infty$ if
\begin{align}
\Rt+\Rh_2^N &> \sum_{k=2}^N \bigl[I(U_k; U^{k-1}, X^N|X_k)+I(X_k; X_2^{k-1})\bigr]+\d(\e'), \label{eq:equivalent}\\
\Rt+\Rh(\hat\Sc^c) &>
\sum_{k \in \hat\Sc^c} H(U_k, X_k) - H(U(\hat\Sc^c), X(\hat\Sc^c)) +3\d(\e') \notag\\
&=\sum_{k\in\hat\Sc^c} \bigl[I(U_k;U(\hat\Sc^c_k), X(\hat\Sc^c)|X_k)+I(X_k; X(\hat\Sc^c_k))\bigr]+3\d(\e'), \quad \hat\Sc\subseteq[2::N], \hat\Sc\ne \emptyset,\label{eq:implied}\\
\Rh(\hat\Sc^c) &> \sum_{k\in\hat\Sc^c} \bigl[I(U_k;U(\hat\Sc^c_k), X(\hat\Sc^c)|X_k)+I(X_k; X(\hat\Sc^c_k))\bigr]+3\d(\e'),\quad \hat\Sc\subseteq[2::N],\hat\Sc\ne[2::N].  \label{eq:imply}
\end{align}
Finally, note that the condition in~\eqref{eq:implied} is inactive
since it is implied by \eqref{eq:imply} and $\Rt>3\d(\e')$.
%


\section{Elimination of Auxiliary Rates in Section~\ref{sec:unicast}}
\label{app:elimination}

To obtain the condition on the message rate $R$ alone, we eliminate the auxiliary rates $\Rh_2, \ldots, \Rh_k$ and $\Rt$ from~\eqref{eq:covering21} through~\eqref{eq:packing2}.
Recalling~\eqref{eq:packing1}, let
\begin{equation}
\Rh_k = I(U_k; Y_k|X_k)-2\d(\e), \quad k \in [2::N].
\label{eq:Rh_k}
\end{equation}
Then substituting \eqref{eq:Rh_k} into~\eqref{eq:covering21} and~\eqref{eq:covering22} yields
\begin{align}
\Rt &> \sum_{k=2}^N \bigl[ I(U_k; U^{k-1}, X^N | X_k, Y_k)+ I(X_k; X_2^{k-1}) \bigr]+2N\d(\e)+\d(\e'), \label{eq:intermediate1}\\
0&> \sum_{k\in\hat\Sc^c} \bigl[ I(U_k; U(\hat\Sc^c_k), X(\hat\Sc^c) | X_k) + I(X_k; X(\hat\Sc^c_k)) - I(U_k; Y_k|X_k)
\bigr]
+2|\hat\Sc^c|\d(\e)+\d(\e') \label{eq:intermediate2}
\end{align}
for all $\hat\Sc\subset[2::N]$. Similarly, substituting \eqref{eq:Rh_k} into~\eqref{eq:packing2} yields
\begin{align}
R+\Rt
&< I(X_1, X(\hat\Sc); U(\hat\Sc^c), Y_N|X(\hat\Sc^c)) \notag\\
&\quad+\sum_{k\in \hat\Sc} \bigl[ I(U_k; U(\hat\Sc_k), U(\hat\Sc^c), X^N|X_k, Y_k)+I(X_k; X(\hat\Sc_k),X(\hat\Sc^c))\bigr] +2(|\hat\Sc|-1)\d(\e)
\label{eq:intermediate3}
\end{align}
for all $\hat\Sc \subseteq [2::N]$ such that $N \in \hat\Sc^c$.
By further eliminating $\Rt$, we obtain the inequalities
\begin{align}
R &< I(X_1, X(\hat\Sc); U(\hat\Sc^c), Y_N|X(\hat\Sc^c))\notag\\
&\qquad-\sum_{k\in \hat\Sc^c} \bigl[I(U_k; U(\hat\Sc^c_k), X^N|X_k, Y_k)
+ I(X_k; X(\hat\Sc^c_k))\bigr] -2|\hat\Sc^c|\d(\e)-\d(\e') \label{eq:ineq4-2}
\end{align}
for all $\hat\Sc\subseteq[2::N]$ such that $N\in\hat\Sc^c$, and
\begin{align}
0&< I(X_1, X(\hat\Sc); U(\hat\Sc^c)|X(\hat\Sc^c)) \notag\\
&\qquad-\sum_{k\in \hat\Sc^c} \bigl[I(U_k; U(\hat\Sc^c_k), X^N|X_k, Y_k)
+I(X_k; X(\hat\Sc^c_k))\bigr] -2|\hat\Sc^c|\d(\e)-\d(\e') \label{eq:ineq4}
\end{align}
for all $\hat\Sc\subset[2::N]$.
Here, \eqref{eq:ineq4-2} follows by combining~\eqref{eq:intermediate1} and~\eqref{eq:intermediate3}, and applying the chain
rule
\begin{align*}
&\sum_{k=2}^N \bigl[ I(U_k; U^{k-1}, X^N | X_k, Y_k) + I(X_k; X_2^{k-1})\bigr] \notag\\
&\qquad= \sum_{k\in \hat\Sc^c} \bigl[ I(U_k; U(\hat\Sc^c_k), X^N|X_k, Y_k)+ I(X_k; X(\hat\Sc^c_k))\bigr] \notag\\
&\qquad\qquad +
\sum_{k\in\hat\Sc} \bigl[ I(U_k;   U(\hat\Sc_k), U(\hat\Sc^c), X^N | X_k, Y_k) + I(X_k; X(\hat\Sc_k),X(\hat\Sc^c))\bigr].
\end{align*}
Inequality~\eqref{eq:ineq4}
follows by rewriting~\eqref{eq:intermediate2} with
\begin{align*}
&\sum_{k\in\hat\Sc^c} \bigl[ I(U_k; U(\hat\Sc^c_k), X(\hat\Sc^c) | X_k) - I(U_k; Y_k|X_k) \bigr]\\
&\qquad=\sum_{k\in\hat\Sc^c} \bigl[ I(U_k; U(\hat\Sc^c_k), X^N | X_k)
    - I(U_k; X_1, X(\hat\Sc)|X(\hat\Sc^c), U(\hat\Sc^c_k))- I(U_k; Y_k|X_k) \bigr]\\
&\qquad=\sum_{k\in\hat\Sc^c} I(U_k; U(\hat\Sc^c_k), X^N | X_k, Y_k)-\sum_{k\in\hat\Sc^c} I(U_k; X_1, X(\hat\Sc)|X(\hat\Sc^c), U(\hat\Sc^c_k))\\
&\qquad=\sum_{k\in\hat\Sc^c} I(U_k; U(\hat\Sc^c_k), X^N | X_k, Y_k)-I(X_1, X(\hat\Sc); U(\hat\Sc^c)|X(\hat\Sc^c)).
\end{align*}
Further rewriting~\eqref{eq:ineq4-2} and~\eqref{eq:ineq4} with $\Sc=\{1\}\cup\hat\Sc$ and $\Sc^c=[1::N]\setminus\Sc$, we have
\begin{align}
R &< I(X(\Sc); U(\Sc^c), Y_N|X(\Sc^c)) \notag\\
&\qquad-\sum_{k\in \Sc^c} \bigl[ I(U_k; U(\Sc_k^c), X^N|X_k, Y_k)
+ I(X_k; X(\Sc^c_k)) \bigr] -2|\Sc^c|\d(\e)-\d(\e'),
\end{align}
for all $\Sc\subseteq[1:N]$ such that $1\in\Sc$, $N\in\Sc^c$ and
\begin{align}
0&< I(X(\Sc); U(\Sc^c)|X(\Sc^c))\notag\\
&\qquad-\sum_{k\in \Sc^c} \bigl[ I(U_k; U(\Sc_k^c), X^N|X_k, Y_k)
+I(X_k; X(\Sc^c_k))\bigr]  -2|\Sc^c|\d(\e)-\d(\e') \label{eq:constraint2}
\end{align}
for all $\Sc\subseteq[1::N]$ such that $1\in\Sc$ and $\Sc^c\ne\emptyset$.
By taking $\e \to 0$ and using the continuity of mutual information,
we can conclude that any rate satisfying
\begin{equation} \label{eq:constraint3}
R < I(X(\Sc); U(\Sc^c), Y_N|X(\Sc^c)) -\sum_{k\in \Sc^c} \bigl[ I(U_k; U(\Sc_k^c), X^N|X_k, Y_k)
+ I(X_k; X(\Sc^c_k)) \bigr]
\end{equation}
for all $\Sc\subseteq[1::N]$ such that $1\in\Sc$, $N\in\Sc^c$ is achievable
under any pmf $p(x^N, u_2^N)$ satisfying
\begin{equation} \label{eq:constraint4}
0 \le I(X(\Sc); U(\Sc^c)|X(\Sc^c))
-\sum_{k\in \Sc^c} \bigl[ I(U_k; U(\Sc_k^c), X^N|X_k, Y_k) +I(X_k; X(\Sc^c_k))\bigr]
\end{equation}
for all $\Sc\subseteq[1::N]$ such that $1\in\Sc$.


\section{Removing the constraint~\eqref{eq:constraint0}} \label{app:constraint-unicast}
We first define some notation that will be used throughout this section.
Let $\Nc=[1::N]$ and for $\Sc \subseteq \Nc$, let
\begin{align}
I(\Sc) &:= I(X(\Sc); Y_N|X(\Sc^c), U(\Sc^c), Q),\label{eq:def-jh} \\
J(\Sc) &:= I(X(\Sc); U(\Sc^c)|X(\Sc^c), Q)\notag\\
&\qquad-\sum_{k\in \Sc^c}\bigl[I(U_k; U(\Sc^c_k), X^N | X_k, Y_k, Q)+I(X_k; X(\Sc^c_k)|Q)\bigr] \label{eq:def-j}\\
&=H(U(\Sc^c), X(\Sc^c)|Q)-\sum_{k\in \Sc^c}\bigl[H(U_k| X_k, Y_k, Q)+H(X_k| Q)\bigr].
\end{align}
Then, \eqref{eq:inequality0} and \eqref{eq:constraint0} can be rewritten as
\begin{align}
R< \max \min_{\substack{\Sc\subseteq\Nc:\\1\in\Sc, N\in\Sc^c}} I(\Sc)+J(\Sc)
\label{eq:rate-K}
\end{align}
where the maximum is over all joint pmfs $p(x^N, u_2^N,q)$ such that
\begin{align}
\label{eq:constraint-K}
J(\Sc)\ge 0, \quad \Sc\subseteq\Nc, 1\in\Sc.
\end{align}
In the following, we will show that the maximum in~\eqref{eq:rate-K} is attained by only considering the distributions that satisfy~\eqref{eq:constraint-K}.

\begin{lemma}\label{lem:alternative-dist}
Let $(X^N, U_2^N, Q)\sim p(x^N, u_2^N, q)$ such that $J(\Ac)<0$ for some $\Ac\subset\Nc$ with $1 \in \Ac$. Then, there exists $(\Xt^N, \Ut_2^N, \Qt) \sim p(\xt^N, \ut_2^N, \qt)$ such that
$\Jt(\Ac) \ge 0$,
\begin{align}
\min_{\substack{\Sc\subseteq\Nc:\\1\in\Sc, N\in\Sc^c}}  I(\Sc) +  J(\Sc)  &< \min_{\substack{\Sc\subseteq\Nc:\\1\in\Sc, N\in\Sc^c}} \It(\Sc) + \Jt(\Sc), \label{eq:ij} \\
\min_{\substack{\Sc\subseteq\Nc:1\in\Sc}}  J(\Sc)  &< \min_{\substack{\Sc\subseteq\Nc:1\in\Sc}} \Jt(\Sc), \label{eq:j1}
\end{align}
where $\It(\Sc)$ and $\Jt(\Sc)$ are~\eqref{eq:def-jh} and~\eqref{eq:def-j} evaluated with $(\Xt^N, \Yt^N, \Ut_2^N, \Qt)$ in place of $(X^N, Y^N, U_2^N, Q)$, respectively, and $\Yt^N$ are the channel output corresponding to the input $\Xt^N$.
\end{lemma}

From Lemma~\ref{lem:alternative-dist}, it follows that
for any achievable rate attained by a distribution with $J(\Ac)<0$ for some $\Ac$,
there exists another distribution such that $\Jt(\Ac) \ge 0$
while strictly increasing the rate and distribution constraints \eqref{eq:ij} and \eqref{eq:j1}. By repeatedly applying Lemma~\ref{lem:alternative-dist} until $J(\Sc) \ge 0$ for all
$\Sc\subseteq\Nc$ such that $1\in\Sc$, we can conclude that
a strictly higher rate is achieved by a distribution
that satisfies the constraint~\eqref{eq:constraint-K}.


It remains to establish Lemma~\ref{lem:alternative-dist}.
We only spell out the proof of \eqref{eq:ij}
since the proof of \eqref{eq:j1} follows essentially the same steps.
To this end, suppose the maximum in~\eqref{eq:rate-K} that is attained by $(X^N, U_2^N, Q)\sim p(x^N, u_2^N, q)$ such that $J(\Ac)<0$ for some $\Ac\subset\Nc$ with $1 \in \Ac$.
If there is no such $\Ac$, then there is nothing to prove.
Let $(\Xt^N, \Ut(\Ac\setminus\{1\}), \Qh)$ be an identically distributed copy of
$(X^N, U(\Ac\setminus\{1\}), Q)$. Let $\Ut(\Ac^c) = \emptyset$ and $\Qt=(\Qh, \Xt(\Ac^c))$.
In other words, $(\Xt^N, \Ut_2^N, \Qt)$ is an identically distributed copy of
$(X^N, U_2^N, Q)$, except that $\Ut(\Ac^c)$ is knocked off and $\Qt$ is augmented with $\Xt(\Ac^c)$.
With this choice of random variables, it is easy to check that $\Jt(\Ac) = 0$.
To establish the inequality in \eqref{eq:ij}, consider
\begin{align}
\min_{\substack{\Sc\subseteq\Nc:\\1\in\Sc, N\in\Sc^c}}
     I(\Sc)+J(\Sc)
&\le\min_{\substack{\Sc\subseteq\Nc:\\1\in\Sc, N\in\Sc^c}}
     I(\Sc\cap\Ac)+J(\Sc\cap\Ac) \label{eq:Sc_cap_Ac}\\
&< \min_{\substack{\Sc\subseteq \Nc:\\1\in\Sc, N\in\Sc^c}}
     \It(\Sc)+\Jt(\Sc), \label{eq:tilde_vars}
\end{align}
where the first inequality holds since $1 \in (\Sc\cap\Ac) \subseteq \Nc$, and the second inequality holds since
\begin{align}
I(\Sc\cap\Ac)
&= I(X(\Sc\cap\Ac); Y_N| X((\Sc\cap\Ac)^c), U((\Sc\cap\Ac)^c), Q)\notag\\
&= I(X(\Sc); Y_N| X(\Ac^c), X(\Sc^c), U(\Ac^c), U(\Sc^c), Q)\notag\\
&\le I(X(\Sc); Y_N| X(\Ac^c), X(\Sc^c), U(\Sc^c), Q)\notag\\
&= \It(\Sc)
\end{align}
and
\begin{align}
J(\Sc\cap\Ac) &= H(U((\Sc\cap\Ac)^c), X((\Sc\cap\Ac)^c)|Q)\notag\\
&\qquad-\sum_{k\in(\Sc\cap\Ac)^c}\bigl[H(U_k|Y_k,X_k,Q)+H(X_k|Q)\bigr]\notag\\
&= H(U((\Sc\cap\Ac)^c), X((\Sc\cap\Ac)^c)|Q)
-H(U(\Ac^c), X(\Ac^c)|Q)\notag\\
&\qquad-\sum_{k\in(\Sc\cap\Ac)^c}\bigl[H(U_k|Y_k,X_k,Q)+H(X_k|Q)\bigr]\notag\\
&\qquad+\sum_{k\in\Ac^c}\bigl[H(U_k|X_k, Y_k,Q)+H(X_k|Q)\bigr]
+J(\Ac)\notag\\
&< H(U((\Sc\cap\Ac)^c), X((\Sc\cap\Ac)^c)|Q)
-H(U(\Ac^c), X(\Ac^c)|Q)\notag\\
&\qquad-\sum_{k\in(\Sc\cap\Ac)^c}\bigl[H(U_k|Y_k,X_k,Q)+H(X_k|Q)\bigr]\notag\\
&\qquad+\sum_{k\in\Ac^c}\bigl[H(U_k|X_k, Y_k,Q)+H(X_k|Q)\bigr] \notag\\
&= H(U(\Sc^c), X(\Sc^c)| X(\Ac^c), U(\Ac^c), Q)\notag\\
&\qquad-\sum_{k\in\Sc^c \setminus \Ac^c}\bigl[H(U_k|Y_k,X_k, Q)+H(X_k|Q)\bigr]\notag\\
&\le H(U(\Sc^c), X(\Sc^c)| X(\Ac^c),  Q)\notag\\
&\qquad-\sum_{k\in\Sc^c \setminus \Ac^c}\bigl[H(U_k|Y_k,X_k,X(\Ac^c), Q)+H(X_k|X(\Ac^c),Q)\bigr]\notag\\
&= H(\Ut(\Sc^c), \Xt(\Sc^c)| \Xt(\Ac^c),  \Qh)\notag\\
&\qquad-\sum_{k\in\Sc^c }\bigl[H(\Ut_k|\Yt_k,\Xt_k,\Xt(\Ac^c), \Qh)+H(\Xt_k|\Xt(\Ac^c),\Qh)\bigr]\notag\\
&= H(\Ut(\Sc^c), \Xt(\Sc^c)| \Qt)-\sum_{k\in\Sc^c }\bigl[H(\Ut_k|\Yt_k,\Xt_k,\Qt)+H(\Xt_k|\Qt)\bigr]\notag\\
&= \Jt(\Sc).\notag
\end{align}


\section{Removing the constraint~\eqref{eq:bc-const1}}\label{app:constraint-broadcast}

We repeat essentially the same argument as in Appendix~\ref{app:constraint-unicast},
so we will be more succinct this time. Define $J(\Sc)$ as in~\eqref{eq:def-j} and $\Nc=[1::N]$.
Then, the rate region characterized by~\eqref{eq:bc1} and~\eqref{eq:bc-const1} can be rewritten as
the rate region that consists of all rate tuples $(R_k \suchthat k \in \Dc)$ such that
\begin{align}\label{eq:ts-bc-rate}
\sum_{k \in \Tc} R_k < \min_{\substack{\Sc\subseteq\Nc:\\1\in\Sc, \Sc^c\cap\Dc=\Tc}} J(\Sc), \quad
\emptyset \ne \Tc \subseteq \Dc,
\end{align}
for some pmf $p(x^N, u_2^N, q)$ such that
\begin{align}\label{eq:ts-bc-const}
\min_{\substack{\Sc\subseteq\Nc:\\1\in\Sc, \Sc^c\cap\Dc = \emptyset}}J(\Sc) \ge 0.
\end{align}
%
%

Following essentially the same steps as those in the proof of Lemma~\ref{lem:alternative-dist},
which is omitted for brevity,
we can show that
the entire rate region~\eqref{eq:ts-bc-rate} is attained by the distributions that satisfy~\eqref{eq:ts-bc-const}.

\begin{lemma}\label{lem:alternative-dist-bc}
Let $(X^N, U_2^N, Q)\sim p(x^N, u_2^N, q)$ such that $J(\Ac)<0$ for some $\Ac\subset\Nc$ with $1 \in \Ac$, $\Ac^c\cap\Dc=\emptyset$. Then, there exists $(\Xh^N, \Ut_2^N, \Qt) \sim p(\xt^N, \ut_2^N, \qt)$ such that
$\Jt(\Ac) \ge 0$,
\begin{align}
\min_{\substack{\Sc\subseteq\Nc:\\1\in\Sc, \Sc^c\cap \Dc=\Tc}}  J(\Sc)  &< \min_{\substack{\Sc\subseteq\Nc:\\1\in\Sc, \Sc^c\cap \Dc=\Tc}} \Jt(\Sc), \quad \emptyset \ne \Tc \subseteq \Dc \label{eq:jbc} \\
\min_{\substack{\Sc\subseteq\Nc:\\1\in\Sc, \Sc^c\cap\Dc=\emptyset}}  J(\Sc)  &< \min_{\substack{\Sc\subseteq\Nc:\\1\in\Sc, \Sc^c\cap\Dc=\emptyset}}  \Jt(\Sc), \label{eq:jbc-const}
\end{align}
where $\Jt(\Sc)$ is~\eqref{eq:def-j} evaluated with $(\Xt^N, \Yt^N, \Ut_2^N, \Qt)$ in place of $(X^N, Y^N, U_2^N, Q)$, and $\Yt^N$ are the output of the channel corresponding to the input $\Xt^N$.
\end{lemma}

From Lemma~\ref{lem:alternative-dist-bc},
it follows that for any achievable rate region attained by a distribution such that there exists some $\Ac$ with $J(\Ac)<0$, there exists another distribution such that $\Jt(\Ac) \ge 0$ while strictly increasing the rate constraints \eqref{eq:jbc} and strictly increasing the constraint on the pmf \eqref{eq:jbc-const}. By repeatedly applying Lemma~\ref{lem:alternative-dist-bc} until $J(\Sc) \ge 0$ for all $\Sc\subseteq\Nc$ such that $1\in\Sc$, $\Sc^c\cap\Dc=\emptyset$, we have shown that there exists a strictly larger achievable rate region which satisfies the constraint~\eqref{eq:ts-bc-const}.


\newcommand{\noopsort}[1]{}

\end{document}